\documentclass[11pt,a4paper,twoside]{article}
\usepackage{graphicx}
\usepackage{amsmath}
\usepackage{amsfonts}
\usepackage{amssymb}
\usepackage{enumerate}
\usepackage{float}
\usepackage[hidelinks]{hyperref}
\usepackage{lscape}
\usepackage{longtable}
\usepackage{mathrsfs}
\usepackage{rotating}
\usepackage{color}
\usepackage{url}
\usepackage{rotating}
\usepackage{yfonts}
\usepackage[labelfont=sl,textfont=sl]{subcaption} % newer version of subfig
\usepackage[font={small,sl},labelsep=period]{caption} % format figure & table captions
\usepackage{titlesec} 
\usepackage[titletoc]{appendix}
\usepackage[ruled,vlined,noline,linesnumbered]{algorithm2e}
\usepackage{tikz}
\usetikzlibrary{automata}
\usetikzlibrary{decorations.markings}
\usetikzlibrary{arrows.meta}
\usetikzlibrary{decorations.pathmorphing}
\newtheorem{theorem}{Theorem}[section]

\newtheorem{corollary}{Corollary}[section]

\newtheorem{definition}{Definition}[section]
\newtheorem{example}{Example}[section]

\newtheorem{lemma}{Lemma}[section]

\newtheorem{proposition}{Proposition}[section]
\newtheorem{remark}{Remark}[section]

\newenvironment{proof}[1][Proof]{\textbf{#1.} }{\hfill$\Box$\vspace{\baselineskip}}
\DeclareMathOperator{\diag}{diag}
\DeclareMathOperator{\spann}{span}

\DeclareMathOperator{\pspan}{pspan}

\newcommand{\T}{\top}
\newcommand{\R}{\mathbb{R}}
\newcommand{\LL}{\mathbb{L}}

\newcommand{\D}{\mathcal{D}}
\newcommand{\mD}{\mathbf{D}}

\DeclareMathOperator{\cols}{col}

\newcommand{\bbm}{\begin{bmatrix}}
\newcommand{\ebm}{\end{bmatrix}}

\newcommand{\vc}{\mathbf{c}}
\newcommand{\ve}{\mathbf{e}}
\newcommand{\vu}{\mathbf{u}}
\newcommand{\vv}{\mathbf{v}}
\newcommand{\vw}{\mathbf{w}}
\newcommand{\vx}{\mathbf{x}}
\newcommand{\vy}{\mathbf{y}}
\newcommand{\vz}{\mathbf{z}}
\newcommand{\zero}{\mathbf{0}}
\newcommand{\zeron}{\mathbf{0}_n}
\newcommand{\one}{\mathbf{1}}
\newcommand{\onen}{\mathbf{1}_n}

\newcommand{\mA}{\mathbf{A}}
\newcommand{\mB}{\mathbf{B}}
\newcommand{\mC}{\mathbf{C}}
\newcommand{\mI}{\mathbf{I}}
\newcommand{\mM}{\mathbf{M}}
\newcommand{\mN}{\mathbf{N}}
\newcommand{\mP}{\mathbf{P}}

\newcommand{\mX}{\mathbf{X}}
\newcommand{\mY}{\mathbf{Y}}

\newcommand{\mPi}{\mathbf{\Pi}}

\newcommand{\calK}{\mathcal{K}}

\newcommand{\rev}[1]{{{\color{black}{#1}}}}

\begin{document}

\title{\centering A characterization of positive spanning sets with ties 
to\\ strongly edge-connected digraphs}
\author{
Denis Cornaz\thanks{LAMSADE, CNRS, Universit\'e Paris Dauphine-PSL, 
Place du Mar\'echal de Lattre de Tassigny, 75016 Paris, France.
(\texttt{denis.cornaz@lamsade.dauphine.fr}).}
\and
S\'ebastien Kerleau\thanks{LAMSADE, CNRS, Universit\'e Paris Dauphine-PSL, 
Place du Mar\'echal de Lattre de Tassigny, 75016 Paris, France 
(\texttt{sebastien.kerleau@lamsade.dauphine.fr}).} 
\and 
Cl\'ement W. Royer\thanks{LAMSADE, CNRS, Universit\'e Paris Dauphine-PSL, 
Place du Mar\'echal de Lattre de Tassigny, 75016 Paris, France. Royer's 
research is partially funded by Agence Nationale de la Recherche through 
program ANR-19-P3IA-0001 (PRAIRIE 3IA Institute). ORCID 0000-0003-2452-2172
(\texttt{clement.royer@lamsade.dauphine.fr}).}
}

\date{\today}

\maketitle

\begin{abstract}
Positive spanning sets (PSSs) are families of vectors that span a given linear 
space through non-negative linear combinations. Despite certain classes of PSSs 
being well understood, a complete characterization of PSSs remains elusive.
In this paper, we explore a relatively understudied relationship between positive 
spanning sets and strongly edge-connected digraphs, in that the former can be viewed 
as a generalization of the latter. We leverage this connection to define a 
decomposition structure for positive spanning sets inspired by the ear 
decomposition from digraph theory. 
\end{abstract}

\paragraph{Keywords} 
Positive spanning sets;
Positive bases; 
Strongly edge-connected digraphs;\\ 
Ear decomposition; 
Network matrices; 
Gaussian elimination.
 
\paragraph{2020 Mathematics Subject Classification} 
% 05C20 Directed graphs (digraphs), tournaments
% 05C50 Graphs and linear algebra
% 05C75 Structural characterization of families of graphs
% 15A03 Vector spaces, linear dependence, rank, lineability
% 15A21 Canonical forms, reduction, classification
% 15B30 Orthogonal matrices
% 15B99 Special matrices (none of the specific categories)
% 90C56 Derivative-free methods and methods using generalized derivatives
05C20; 05C50; 15A21; 15B99.

%%%%%%%%%%%%%%%%%%%%%%%%%%%%%%%%%%%%%%%%%%%%%%%%%%%%%%%%%%%%%%%%%%%%%%%%%%%%%%%
\section{Introduction}
\label{sec:intro}
%%%%%%%%%%%%%%%%%%%%%%%%%%%%%%%%%%%%%%%%%%%%%%%%%%%%%%%%%%%%%%%%%%%%%%%%%%%%%%%

Gaussian elimination is a fundamental technique in linear algebra, that can be 
used to assess whether a given matrix is linearly spanning, in the sense that 
its columns span the entire space through linear 
combinations~\cite[Page 6]{JLee_2022}. Similarly, in graph theory, one can  
determine a spanning tree of a graph using linear algebra 
techniques, while efficient implementations of graph algorithms can be 
obtained by leveraging sparse linear algebra~\cite{JTFineman_ERobinson_2011}.

Positive spanning sets, or PSSs, are matrices such that the columns span 
the entire space through nonnegative linear combinations~\cite{CDavis_1954}. 
These matrices are instrumental to direct-search algorithms, a 
class of continuous optimization algorithms that proceed by exploring the 
variable space through suitably chosen 
directions~\cite{TGKolda_RMLewis_VTorczon_2003,CAudet_WHare_2017}.  When 
those directions are chosen from positive spanning sets, convergence can 
be guaranteed at a rate that heavily depends on the properties of the PSSs 
at hand~\cite{TGKolda_RMLewis_VTorczon_2003,
MDodangeh_LNVicente_ZZhang_2016}. In this setting, using a direction 
corresponds to evaluating an expensive function, and thus optimizers 
typically rely on inclusion-wise minimal positive spanning sets, or 
positive bases~\cite{RMcKinney_1962,JRReay_1966}. Although positive bases 
have already been fully described~\cite{ZRomanowicz_1987}, generic 
\rev{descriptions} are often impractical to generate positive bases in practice. 
As a result, optimizers have focused on characterizing special positive 
bases for which simpler \rev{characterizations} can be 
obtained~\cite{WHare_GJarryBolduc_SKerleau_CWRoyer_2024,
WHare_GJarryBolduc_CPlaniden_2023}.

Perhaps surprisingly, a connection between positive spanning sets and strongly 
edge-connected digraphs was spotted early in the PSS 
literature~\cite{DAMarcus_1981}, but to the authors' knowledge this connection has 
not been exploited further.
%\footnote{More precisely, Marcus~\cite{DAMarcus_1981} claims that this connection will be investigated in a future work that we believe has never been released.}. 
Meanwhile, numerous results have been established 
for strongly connected digraphs~\cite{JBangJensen_GZGutin_2009,
SKhuller_BRaghavachari_NYoung_1996}, with minimal strongly edge-connected 
digraphs attracting recent interest~\cite{MArcosArgudo_JGarciaLopez_LMPozoCoronado_2019,
DCornaz_HKerivin_ARMahjoub_2018,JGarciaLopez_CMarijuan_2012,
JGarciaLopez_CMarijuan_LMPozoCoronado_2018}. Although such digraphs appear 
connected to positive bases through the concept of minimality, a formal link 
between those objects has yet \rev{to} be described.

In this paper, we provide certificates for the positive spanning property 
based on digraph theory. To this end, we show that PSSs can be seen as 
generalizing the concept of strongly edge-connected digraphs. We then 
leverage this connection to obtain a novel characterization of such matrices 
based on the ear decomposition of digraphs~\cite{ASchrijver_2003}.  

% Structure
The remainder of this paper is organized as follows. In 
Section~\ref{sec:digraphs}, we review key results from digraph theory. We then 
discuss positive spanning sets and draw connections with strongly edge-connected 
digraphs in Section~\ref{sec:pss}. Our main results, that generalize the ear 
decomposition to positive spanning sets, are derived in 
Section~\ref{sec:eardec}.

\paragraph{Notations}
Throughout this paper, we work in the Euclidean space $\R^n$ with $n \ge 2$, 
or a linear subspace thereof, denoted by $\LL \subset \R^n$. The dimension 
of such a subspace will always be assumed to be at least $1$.
The set of real matrices with $n$ rows and $m$ columns will be denoted as 
$\R^{n\times m}$. Those dimensions will always be assumed to be at least $1$.
Bold lowercase letters (e.g. $\mathbf{v}$,$\mathbf{a}$) will be used to 
designate vectors and arcs in directed graphs, while bold uppercase letters 
(e.g. $\mathbf{D}$) will denote matrices. 
The notations $\zeron$ and $\onen$ will respectively be used to designate the 
null vector and the all-ones vector in $\R^n$, while 
$\mathbf{I_n}=\bbm\mathbf{e_1}&\dots&\mathbf{e_n}\ebm$  will denote the 
identity matrix in $\R^{n \times n}$.
Given a matrix $\mathbf{D} \in \R^{n\times m}$, its set of columns will be 
denoted $\cols(\mD)$ while its linear span ($i.e.$ the set of linear 
combinations of its columns) will be denoted by $\spann(\mathbf{D})$. 
The matrix whose entries are the signs of those of $\mM$ will be noted 
$sgn(\mM)$. 
Calligraphic letters such as $\D$ and $\mathcal{S}$ will be used for finite 
families of vectors or of indices. 
\rev{For any integer $m \ge 1$, we let 
$[\![1,m]\!]:=\{z: 1\leq z \leq m, z \in \mathbb{Z}\}$.}
Finally, for a digraph $G=(V,A)$ the notations 
$(u,v)$ and $u-v$ will respectively designate an arc and an oriented path 
in $A$ going from $u$ to $v$.

%%%%%%%%%%%%%%%%%%%%%%%%%%%%%%%%%%%%%%%%%%%%%%%%%%%%%%%%%%%%%%%%%%%%%%%%%%%%%%%
\section{Digraphs and ear decomposition}
\label{sec:digraphs}
%%%%%%%%%%%%%%%%%%%%%%%%%%%%%%%%%%%%%%%%%%%%%%%%%%%%%%%%%%%%%%%%%%%%%%%%%%%%%%%

In this section, we recall classical results on the ear decomposition for 
digraphs. For sake of completeness, we first define the main concepts and 
properties of digraphs to be used throughout the 
paper~\cite{JBangJensen_GZGutin_2009,ASchrijver_2003}.
We consider digraphs of the form $G=(V,A)$, where $V$ denotes a set of 
vertices and $A$ denotes a set of arcs. 
A \emph{directed path} in $G$ is a sequence of arcs of 
the form $\{(u_i,u_{i+1})\}_{i=1,\dots,k}$. Directed paths such that 
$u_1=u_{k+1}$ with $u_2, \dots, u_k$ all distinct are called \emph{circuits}. 
A digraph $(V,A)$ is called \emph{acyclic} if the set $A$ does not contain 
any circuit, while it is called \emph{strongly edge-connected} - or 
\emph{strongly connected} for simplicity- if for any $(u,v) \in V^2$, 
there exists a directed path from $u$ to $v$. 
A strongly connected digraph $G=(V,A)$ is \emph{minimally strongly connected} 
if any digraph $G'=(V,A')$ such that $A' \subset A$, $|A'|=|A|-1$ is not 
strongly connected.
Finally, an \emph{oriented spanning tree} of a graph $G$ is an oriented tree 
$T=(V,\hat{A})$ such that $\hat{A}\subset A$.

As mentioned in the previous section, we are interested in certifying whether 
a digraph is strongly connected. Proposition~\ref{prop:cutstrongconnected}
provides a negative certificate for this property.

\begin{proposition}
\label{prop:cutstrongconnected}
    A connected digraph $G=(V,A)$ is \emph{not} strongly edge-connected 
	if and only if there exists an \emph{oriented cut} of $G$, i.e. a set 
    $\tilde{A}\subset A$ and two vertex-disjoint subgraphs $G_1=(V_1,A_1)$ 
    and $G_2=(V_2,A_2)$ of $G$ with $V_1\neq \emptyset$, 
    $V_2 \neq \emptyset$, such that
	\[
        A=\tilde{A}\cup A_1\cup A_2 
        \quad \text{and} \quad
		\forall (u,v) \in \tilde{A}, 
		\quad 
		u \in V_1\ \mbox{and}\ v \in V_2.
	\]	
\end{proposition}
\begin{proof}
	Suppose first that there exists an oriented cut of $G$ given by 
    $\tilde{A} \subset A$, $G_1=(V_1,A_1)$ and $G_2=(V_2,A_2)$. 
    Let $v_1 \in V_1$ and $v_2 \in V_2$. 
	By definition of an oriented cut, there does not exist a directed path in $A$ 
	joining $v_2$ to $v_1$, proving that $G$ is not strongly connected.
	
	Conversely, suppose that $G$ is not strongly connected and let $v_1,v_2$ 
	be two vertices for which there is no $v_2-v_1$ path in $A$. 
    Let $V_2\subset V$ be the \rev{set} of vertices $v$ such that $A$ contains a 
    $v_2-v$ path and let $V_1 = V \setminus V_2$. 
    If $A_1$ (resp. $A_2$) denotes the set of arcs between 
	vertices of $V_1$ (resp. $V_2$), then $G_1=(V_1,A_1)$, $G_2=(V_2,A_2)$ and 
    $\tilde{A}=A\backslash\left(A_1\cup A_2\right)$ define an oriented cut for 
    $G$ as by construction, the arcs in 
	$\tilde{A}$ must be of the form $(u_1,u_2)$ with 
	$u_1 \in V_1$ and $u_2 \in V_2$.
\end{proof}

We now turn to providing a positive certificate for strongly connected 
digraphs, based on the concept of ear 
decomposition~\cite[Section 5.3]{JBangJensen_GZGutin_2009}.

\begin{definition}[Ear and ear decomposition]
\label{def:ear}
	Let $G=(V,A)$ be a digraph. An \emph{ear} of $G$ is a directed path 
	$\{(u_i,u_{i+1})\}_{i=1,\dots,k} \subset A$ such that for any 
	$i \in [\![2,k]\!]$, $u_i$ is the head and the tail of exactly 
	one arc in $A$. 
	
	The graph $G$ possesses an \emph{ear decomposition} if there 
	exists a sequence of digraphs $\{G_i=(V_i,A_i)\}_{i=1,\dots,s}$ such that 
	\begin{enumerate}[(i)]
		\item $V_1 \subseteq V_2 \subseteq \cdots \subseteq V_s=V$,
		\item $A_1 \subseteq A_2 \subseteq \cdots \subseteq A_s=A$,
		\item $G_1$ consists of one vertex and no arcs,
		\item For any $i \in  [\![2,s]\!]$, $A_i\backslash A_{i-1}$ defines an 
        ear of $G_{i}$.
	\end{enumerate}	
\end{definition}

\begin{figure}[h!]
\centering
\begin{tikzpicture}
\tikzstyle{vertex}=[circle, fill=white, minimum size=7pt, inner sep=1pt]
\tikzset{->-/.style={decoration={
  markings,
  mark=at position .5 with {\arrow{>}}},postaction={decorate}}}
\tikzstyle{edge}=[->-, very thick]
\node[vertex] (v1) at (0,0) {$v_1$};
\node[vertex] (v2) at (2,0) {$\textcolor{blue}{v_2}$};
\node[vertex] (v3) at (4,0) {$\textcolor{purple}{v_6}$};
\node[vertex] (v4) at (0,2) {$\textcolor{blue}{v_4}$};
\node[vertex] (v5) at (2,2) {$\textcolor{blue}{v_3}$};
\node[vertex] (v6) at (4,2) {$\textcolor{purple}{v_5}$};
\draw[edge,blue] (v1) to (v2);
\draw[edge,blue] (v2) to (v5);
\draw[edge,blue] (v5) to (v4);
\draw[edge,blue] (v4) to (v1);
\draw[edge,purple] (v2) to (v3);
\draw[edge,purple] (v3) to (v6);
\draw[edge,purple] (v6) to (v5);
\end{tikzpicture}
\caption{An ear decomposition $(G_1,G_2,G_3)$ where $G_1=(\{v_1\},\emptyset)$, 
$G_2$ is obtained by adding the blue arcs and vertices and $G_3$ is obtained by 
adding the red arcs and vertices.}
\label{fig:eardecomp}
\end{figure}

Figure~\ref{fig:eardecomp} below provides an example of a strongly connected 
graph that possesses an ear decomposition. In fact, ear decompositions 
characterize strongly connected digraphs in the following sense.

\begin{theorem}~\cite[Theorem  6.9]{ASchrijver_2003}
\label{th:eardecompograph}
	A digraph $G$ is strongly edge-connected if and only if it possesses an ear 
	decomposition.
\end{theorem}
Theorem~\ref{th:eardecompograph} thus provides a positive certificate for the 
strongly connected property. In this paper, we aim at generalizing the result 
of Theorem~\ref{th:eardecompograph} to positive spanning sets, which we define 
in the next section.

%%%%%%%%%%%%%%%%%%%%%%%%%%%%%%%%%%%%%%%%%%%%%%%%%%%%%%%%%%%%%%%%%%%%%%%%%%%%%%%
\section{Positive spanning sets}
\label{sec:pss}
%%%%%%%%%%%%%%%%%%%%%%%%%%%%%%%%%%%%%%%%%%%%%%%%%%%%%%%%%%%%%%%%%%%%%%%%%%%%%%%

The notion of positive spanning set (PSS) is a classical concept from linear 
algebra. The first part of this section reviews classical results on PSSs, 
while the second part draws connections between PSSs and strongly connected 
digraphs.

%%%%%%%%%%%%%%%%%%%%%%%%%%%%%%%%%%%%%%%%%%%%%%%%%%%%%%%%%%%%%%%%%%%%%%%%%%%%%%
\subsection{Definition and characterization}
\label{ssec:pss}

Positive spanning sets are commonly defined as families of vectors, akin to 
spanning sets. In this paper, we adopt the following, equivalent definition 
based on matrices.

\begin{definition}[Positive span and positive spanning set]
\label{def:pspanpss}
	Let $\LL$ be a linear subspace of $\R^n$ and $m \ge 1$. The 
	\emph{positive span} of a matrix $\mD \in \R^{n\times m}$, 
	denoted by $\pspan(\mD)$, is the set 
	$$
		\pspan(\mD):=\{\mD\vx\,|\, \vx \in \R^m, \vx \geq \zero_m\}.
	$$
	A \emph{positive spanning set} (PSS) of $\LL$  is a matrix $\mD$ 
	such that $\pspan(\mD)=\LL$. When $\LL=\R^n$, the matrix $\mD$ will simply 
	be called a positive spanning set.
\end{definition}

Several characterizations of positive spanning sets have been proposed in the 
literature. Proposition~\ref{prop:characterizepss} summarizes those that are 
relevant for this paper~\cite{WHare_GJarryBolduc_SKerleau_CWRoyer_2024,
RGRegis_2016}.

\begin{proposition}
\label{prop:characterizepss}
	Let $\mD \in \R^{n\times m}$. 
	The following statements are equivalent.
	\begin{enumerate}[(i)]
		\item $\mD$ is a PSS for some linear subspace $\LL \subset \R^n$.
		\item $\pspan(\mD)=\spann(\mD)$.
		\item\label{item3:propcharactPSS} There exists a \emph{positive} 
		vector $\vx \in \R^m$ such that $\mD\vx=\zero_n$.
	\end{enumerate}
\end{proposition}
When the matrix $\mD$ has full rank, statement \eqref{item3:propcharactPSS} of 
Proposition~\ref{prop:characterizepss} provides a certificate that $\mD$ is a 
PSS of $\R^n$. For future reference, we now present a certificate that a matrix 
\emph{does not} positively span the entire space, which is a variant of Farkas' 
lemma~\cite{ADax_1997}.

\begin{proposition}\rev{\cite[Theorem 2.3]{ARConn_KScheinberg_LNVicente_2009b}}
\label{prop:certifnopss}
	A matrix $\mD \in \R^{n \times m}$ does \emph{not} positively span $\R^n$ if 
	and only if there exists a non-zero vector $\vy \in \R^n$ such that 
	$\vy^\top\mD \geq\zero_m^\top$.
\end{proposition}

The positive spanning property is invariant to several operations on matrix 
columns, such as rescaling or permutation. In addition, if $\mD$ is a PSS for 
a given $\ell$-dimensional space, then for any invertible matrix $\mB$, the 
matrix $\mB^{-1}\mD$ is a PSS for another $\ell$-dimensional space. 
These invariance properties imply that we can reduce the study of 
positive spanning sets to equivalent classes defined as follows.

\begin{definition}[Structural equivalence]~\label{def:structequiv}
	Let $\mD$ and $\mD^{'}$ be two matrices in $\R^{n \times m}$. The matrices 
	$\mD$ and $\mD^{'}$ are \emph{structurally equivalent} if there exists
	a non-singular matrix $\mB \in \R^{n \times n}$, a permutation matrix 
	$\mP \in \R^{m \times m}$ and a diagonal matrix 
	$\mathbf{\Delta} \in \R^{m \times m}$ with positive diagonal entries such 
	that
	\[
		\mD^{'} \; = \; \mB^{-1} \mD \mP \mathbf{\Delta}.
	\] 
	We then write $\mD \equiv \mD^{'}$.
\end{definition}

Note that the notion of structural equivalence for PSSs was previously 
stated in the context of derivative-free optimization for families of 
vectors~\cite[Definition 2.3]{IDCoope_CJPrice_2001}. By adapting this 
definition to matrices, we can combine Definition~\ref{def:structequiv} 
together with Proposition~\ref{prop:characterizepss} to obtain a 
characterization of PSSs based on structural equivalence. \rev{This characterization 
involves matrices that have a particularly simple expression, thanks to the rescaling 
and permutation operators.}

\begin{proposition}\label{prop:nicestructurepss}
	A matrix $\mD \in \R^{n \times m}$ is a PSS of some  $\ell$-dimensional subspace $\LL$ of $\R^n$ if and only if there exists $m-n$ vectors 
	$\vv_1$,$\dots$,$\vv_{m-n}$ in $\R^{\ell}$ such that
	\begin{equation}
	\label{eq:nicestructurepss}
		\mD 
		\; \equiv \; 
		\bbm 
		\mI_{\ell} & \vv_1 &\cdots &\vv_{m-n} \\
		\zero_{n-\ell,\ell} &\zero_{n-\ell} &\cdots &\zero_{n-\ell} 
		\ebm
		\quad \mbox{and} \quad
		\sum_{i=1}^{m-n} \vv_{i} = - \one_{\ell}.
	\end{equation}
	In particular, if $m=\ell+1$, one has
	\begin{equation}
	\label{eq:nicestructureminpb}
		\mD 
		\; \equiv \;
		\bbm
			\mI_{\ell} &-\one_{\ell} \\
			\zero_{n-\ell,\ell} &\zero_{n-\ell}
		\ebm
		.
	\end{equation}
\end{proposition}

The second part of Proposition~\ref{prop:nicestructurepss} provides a 
characterization of a subclass of PSSs called minimal positive bases (see 
Section~\ref{sec:apps} for a formal definition). However, for general 
matrices, the structural equivalence~\eqref{eq:nicestructurepss} is not 
satisfactory, as it does not provide an easy certificate for verifying whether 
a matrix is a PSS of a given subspace.

%%%%%%%%%%%%%%%%%%%%%%%%%%%%%%%%%%%%%%%%%%%%%%%%%%%%%%%%%%%%%%%%%%%%%%%%%%%%%%%
\subsection{Connection with digraphs}
\label{ssec:network}

Having defined key concepts associated with PSSs, we now formalize their 
relationship with (strongly connected) digraphs. Although existing connections 
involve incidence matrices~\cite{DAMarcus_1981}, our results rely more 
generally on network matrices~\cite{ASchrijver_2003}.

\begin{definition}[Network matrix]
\label{def:networkmat}
	Let $G=(V,A)$ be a digraph with $A=\{(u_j,v_j)\}_{j=1}^m$ and let 
	$T=(V,\hat{A})$ be an oriented spanning tree with 
	$\hat{A} =\{(\hat{u}_i,\hat{v}_i)\}_{i=1}^n$. The network 
	matrix associated with $G$ and $T$ is the matrix $\mM \in \R^{n \times m}$ 
	defined by 
	\[
		\forall i \in [\![1,n]\!],\ 
		\forall j \in [\![1,m]\!],
		\quad
		\mM_{i,j}= \left\{ 
			\begin{array}{ll}
                0 	&\text{if the path $u_j-v_j$ in $T$ does not pass through 
                    $(\hat{u}_i,\hat{v}_i)$,}\\
				1  	&\text{if $u_j-v_j$ passes through $(\hat{u}_i,\hat{v}_i)$ 
					in forward direction,} \\
				-1 	&\text{if $u_j-v_j$ passes through $(\hat{u}_i,\hat{v}_i)$ 
					in backward direction.}
			\end{array}
		\right. 
	\] 
\end{definition}

\rev{An example of network matrix is provided in 
Figure~\ref{fig:digraph and netwmat}.}
\begin{figure}[H]
    \centering
    \begin{minipage}{0.45\textwidth}
    \centering
    \begin{tikzpicture}
        \tikzstyle{vertex}=[circle, fill=black, minimum size=7pt, inner sep=1pt]
        \tikzset{->-/.style={decoration={
        markings,
        mark=at position .5 with {\arrow{>}}},postaction={decorate}}}
        \tikzstyle{edge}=[->-, very thick]
        \tikzstyle{edge2}=[->-, dashed, very thick]
        \node[vertex] (v1) at (0,0) {};
        \node[vertex] (v2) at (2,0) {};
        \node[vertex] (v3) at (4,0) {};
        \node[vertex] (v4) at (2,2) {};
        \node at (1.1,0.35) {1};
        \node at (3.1,0.35) {2};
        \node at (1.65,0.9) {4};
        \node at (1.9,-0.55) {3};
        %\node at (3.2,1.2) {5};
        \draw[edge] (v2) to (v1);
        \draw[edge] (v3) to (v2);
        \draw[edge] (v2) to (v4) ;
        %\draw[edge2] (v4) to (v3);
        \draw[edge2] (v1) to [bend right=45] (v3);
    \end{tikzpicture}
    \end{minipage}%
    \hfill
    \begin{minipage}{0.55\textwidth}
    \centering
    $\bbm1&0&-1&0\\
       0&1&-1&0\\
       0&0&0&1\\
    \ebm$
    \end{minipage}
    \caption{\rev{Digraph, spanning tree (plain edges) and network matrix. The arcs are 
    numbered to match the column ordering}}
    \label{fig:digraph and netwmat}
\end{figure}

\rev{Several remarks are in order regarding Definition 3.3.}
First, note that there is no ambiguity in the definition as any path joining two 
vertices in $T$ is unique. Secondly, notice that we arbitrarily selected an 
ordering of the arcs in both $A$ and $\hat{A}$. Two network matrices defined 
using a different ordering of those arcs are structurally equivalent in the 
sense of Definition~\ref{def:structequiv}
\rev{\cite[(36) p.277]{ASchrijver_2003}}. In the rest of the paper, unless 
otherwise needed, we will not specify the arc ordering.
Finally, one may consider the incidence matrix of a digraph $G=(V,A)$ as a 
network matrix associated with $G$ and with the spanning tree 
$T=(V \cup \{u\},\{(u,v)\}_{v \in V})$. 

In this paper, for simplicity, we focus 
on network matrices whose associated spanning tree $T$ is a subgraph of the 
associated digraph $G$, as such matrices are structurally equivalent to one 
another. For sake of completeness, we state and prove this property in the 
proposition below.

\begin{proposition}
\label{prop: network matrices all equiv}
Let $G=(V,A)$ be a strongly connected digraph with $|V|=n+1$ and $|A|=m$. Let 
$T=(V,\hat{A})$ with $\hat{A}\subset A$ be a spanning tree of $G$ and let 
$\mM \in \R^{n\times m}$ be the associated network matrix. 
%The following statements are equivalent.
%\begin{itemize}
%    \item[i)] $\mM' \in \R^{n \times m}$ is a network matrix for $G$ and for 
%    some spanning tree $T'=(V,\hat{A}')$, $\hat{A}'\subset A$.
%    \item[ii)] $\mM'=\mathbf{B}^{-1}\mathbf{MP}$ where $\mB\in\R^{n\times n}$ 
%    is nonsingular, $\cols(\mB) \subset \cols(\mM)$ and $\mP\in \R^{m\times m}$ is a 
%    permutation matrix.
%\end{itemize}
Then, a matrix $\mM'$ is a network matrix for $G$ and a spanning tree 
$T'$ if and only if $\mM'=\mathbf{B}^{-1}\mathbf{MP}$ where 
$\mB\in\R^{n\times n}$ is nonsingular, $\cols(\mB) \subset \cols(\mM)$ and 
$\mP\in \R^{m\times m}$ is a permutation matrix.
\end{proposition}

\def\oldproofProp{0}

\begin{proof}
Letting $A=\{a_1,\dots,a_m\}$ and $T=(V,\hat{A})$, we assume without loss 
of generality that $\hat{A}=\{a_1,\dots,a_n\}$ and that the columns of $\mM$, 
denoted by $\vc_1,\dots,\vc_m$, correspond to the arcs $a_1,\dots,a_m$ in that 
order. Since $\mM$ is a network matrix, we thus have $\vc_i=\ve_i$ for every 
$i \in \{1,\dots,n\}$, while any $\vc_i$ with $i \in \{m+1,\dots,n\}$ has 
coefficients in $\{-1,0,1\}$.

%\textbf{Forward implication} 
Suppose first that $\mM'$ is a network matrix 
associated with $G$ and a spanning tree $T'=(V,\hat{A}')$ with 
$\hat{A}' \subset A$, and let $a_{i_1},\dots,a_{i_n}$ denote the arcs in $\hat{A}'$ 
with $1 \le i_1 < i_2 < \cdots < i_n \le m$. There exists a permutation matrix 
$\mP \in \R^{m \times m}$ such that the columns of $\mPi=\mM'\,\mP^{-1}$, 
denoted by $\vc_1',\dots,\vc_m'$, correspond to arcs $a_1,\dots,a_m$ in that 
order. Then, for any $1 \le j \le n$, we have $\vc_{i_j}'=\ve_j$. Moreover, since 
$\hat{A}'$ defines a tree on $G$, the columns $\vc_{i_1},\dots,\vc_{i_n}$ of $\mM$ 
define a basis of $\R^n$. Letting $\mB \in \R^{n \times n}$ be the matrix defined 
with those columns in order, it follows that $\mB$ is a nonsingular matrix defining 
the change of basis from $\{\vc_i\}_{i=1,\dots,n}$ to 
$\cols(\mB)=\{\vc_{i_j}\}_{j=1,\dots,n}$. As a result, we have 
$\mB\vc_{i_j}'=\mB\ve_j=\vc_{i_j}$ for any $1 \le j \le n$ and $\mB\vc_i'=\ve_i$ 
for any $1 \le i \le n$, from which it follows that $\mB\mPi=\mM$.
Overall, we have shown that $\mB\mM'\mP^{-1}=\mM$ with $\mB$ nonsingular such 
that $\cols(\mB) \subset \cols(\mM)$ and $\mP$ a permutation matrix.

%\textbf{Reverse implication} 
Suppose now that $\mM'=\mB^{-1}\mM\mP$ with 
$\mB \in \R^{n \times n}$ nonsingular such that 
$\cols(\mB) \subset \cols(\mM)$ and $\mP$ a permutation matrix. Since $\mB$ is 
nonsingular, the columns of $\mB$, which we denote by $\vc_{i_1},\dots,\vc_{i_n}$ 
with $1 \le i_1 < i_2 < \cdots < i_n \le m$, define a basis of $\R^n$. 
Consequently, those columns define a set of arcs 
$\hat{A}'=\{a_{i_j}\}_{j=1,\dots,n}$ such that $T'=(V,\hat{A}')$ is a spanning tree 
for $G$. Since $\mB$ represents the change of basis from $\vc_1,\dots,\vc_n$ to 
$\vc_{i_1},\dots,\vc_{i_n}$, the columns of $\mB^{-1}\mM$ represent the expression 
of each path in the basis $\{\vc_{i_j}\}_{j=1,\dots,n}$. It follows that 
$\mB^{-1}\mM=\mM'\mP^{-1}$ is a network matrix, hence $\mM'$ is also a network 
matrix associated with the same spanning tree.
\end{proof}

\if\oldproofProp

\textcolor{blue}{OLD PROOF BELOW}

Let $|V|=n$, let  with $m\geq n$ and for the sake of simplicity assume that $\hat{A}=\{a_1,\dots,a_{n-1}\}$. For all $i,$ note $\mathbf{c}_i$ the column of $\mM$ associated to $a_i$ and for the sake of simplicity assume that it is of index $i$.\\[0.25cm] 

First, let $\mM'$ be a network matrix associated to $G$ and to some spanning tree $T'=(V,\hat{A}')$, satisfying $\hat{A}'=\hat{A}\backslash\{a_k\}\cup\{a_\ell\}$, where $k\leq n-1$ and $\ell\geq n$: in other words, $\hat{A}$ and $\hat{A}'$ differ from one another by exactly one arc. Moreover, assume that for all $i$, the arc $a_i=(u_i,v_i) \in A$ is associated to the $i^{th}$ column of $\mM'$, namely $\mathbf{c}'_i$. As a consequence, $\mathbf{c}_i=\mathbf{c}'_i=\mathbf{e}_i$ whenever $i\leq n-1, i\neq k$. Moreover $\mathbf{c}_k=\mathbf{c}'_\ell=\mathbf{e}_k$.\\
For all $i$, the column $\mathbf{c}_i$ (resp. $\mathbf{c}_i'$) represents the unique $u_i-v_i$ path $p_i$ in $T$ (resp. $p_i'$ in $T'$). The two paths $p_i$ and $p_i'$ are equal on the sole condition that $p_i$ does not use $a_k$. More generally, letting $a_k=(u_k,v_k)$, $p_i'$ is the path that
\begin{itemize}
    \item[i)] follows $p_i$ until reaching either $u_k$, $v_k$ or $v_i$,
    \item[ii)] if $u_k$ or $v_k$ was reached, follows the path $p'_k$ path in $T'$ from end to end,
    \item[iii)] keeps following $p_i$ to its end.
\end{itemize}
Note that the path $p'_i$ described by this procedure can sometimes go through a same arc on two distinct occasions. In such cases, the two crossings happen in opposite directions and cancel each other: the arc is considered not to be part of $p_i'$.\\
Based on these observations, we find that for any arc $a_i=(u_i,v_i) \in A$ with $\mathbf{c}_i=\sum\limits_{j=1}^{n-1}\mu_j\mathbf{c_j}$, we must have $\mathbf{c_i}'=\sum\limits_{\substack{j=1 \\ j \neq k}}^{n-1}\mu_j\mathbf{c}'_j + \mu_k\left(\sum\limits_{\substack{j=1 \\ j \neq k}}^{n-1}\lambda_j\mathbf{c}'_j+\lambda_\ell\mathbf{c}'_\ell\right)$ where $\mathbf{c}'_k=\sum\limits_{\substack{j=1 \\ j \neq k}}^{n-1}\lambda_j\mathbf{c}'_j+\lambda_\ell\mathbf{c}'_\ell$ (and so $\lambda_\ell \neq 0).$\\ 
In other words $\mathbf{c_i}'=\sum\limits_{\substack{j=1 \\ j \neq k}}^{n-1}(\mu_j+\mu_k\lambda_j)\mathbf{c}'_j+\mu_k\lambda_\ell\mathbf{c}'_{\ell}$. Now, since $T'$ is obtained by replacing the arc $a_k \in \hat{A}$ with $a_\ell \in \hat{A}'$, for all $j\leq n-1$ with $j\leq k$, $\mathbf{c}'_j=\mathbf{c}_j$ and $\mathbf{c}'_\ell=\mathbf{c}_k=\mathbf{e}_k$. 
Thus $\mathbf{c}'_i=\sum\limits_{\substack{j=1 \\ j \neq k}}^{n-1}(\mu_j+\mu_k\lambda_j)\mathbf{c}_j+\mu_k\lambda_\ell\mathbf{c}_{k}$.\\
Let $\mB'$ be the matrix obtained from $\mI_{n-1}$ by replacing its $k^{th}$ column with $\bbm \lambda_1&\dots&\lambda_{n-1}\ebm^\top$. Since $\lambda_k\neq 0$ the matrix $\mB'$ is invertible. Moreover, by what precedes $\mB'\mathbf{c}_i=\mathbf{c}'_i$, for all $i$. Therefore, letting $\mB=\mB'^{-1}$ we obtain $\mB^{-1}\mM=\mM'$. Since $\mM'$ contains an identity block, it is clear that $\cols(\mB)\subset \cols(\mM')$ and the result is proved. In the general case, note that two columns of $\mM$ and $\mM'$ with the same index might not always be associated to the same arc. However, such scenarios are easily dealt with through a multiplication by a permutation matrix $\mP$. Finally, the result is easily extended to any spanning tree $T'$ and any associated network matrix $\mM'$ as for any two spanning trees $T$ and $T'$ one can always find a finite sequence of trees, each differing from the previous by exactly one arc, of first term $T$ and of last term $T'$.
\\[0.25cm]
Conversely, let $\mB\in \R^{(n-1)\times(n-1)}$ be nonsingular with $\cols(\mB) \subset \cols(\mM)$ and let $\mM'=\mB^{-1}\mM\mP$ where $\mP$ is a permutation matrix. Clearly, the canonical basis of $\R^{n-1}$ is contained in $\cols(\mM')$ and proving that $\mM'$ is a network matrix associated to $G$ and to some spanning tree $T'$ is equivalent to proving the same result for $\mB^{-1}\mM\mP'$, for any other permutation matrix $\mP'$. Therefore, for the sake of simplicity, suppose that 
$\mP=\mI_{n-1}$ and note $\mM'=\mB^{-1}\mM=\bbm\mathbf{c}_1'&\dots&\mathbf{c}_{m}'\ebm$. The first $n-1$ columns of $\mM=\bbm\mathbf{c}_1&\dots&\mathbf{c}_m\ebm$ form the block $\mI_{n-1}$ and at first, we suppose that $\cols(\mB)=\{\mathbf{c}_1,\dots,\mathbf{c}_{n-1}\}\backslash\{\mathbf{c}_k\}\cup\{\mathbf{c}_\ell\}$ for some $k\leq n-1$ and $\ell\geq n.$ As $\mB$ is nonsingular, letting $\mathbf{c}_\ell=\sum\limits_{j=1}^{n-1}\alpha_j\mathbf{c}_j$, we must have $\alpha_k \in \{-1,1\}$. As a consequence, letting $a_\ell=(u_\ell,v_\ell) \in A$ be associated to $\mathbf{c}_\ell$, the unique $u_\ell-v_\ell$ path in $T$, named $p_\ell$, uses $a_k$.\\
Now, letting $\hat{A}'=\{a_1,\dots,a_{n-1}\}\backslash\{a_k\}\cup\{a_\ell\}$ be the set of arcs associated to $\cols(\mB)$, we see that $|\hat{A}'|=n-1$. Since $T=(V,\hat{A})$ is a tree and $\hat{A}'\backslash\{a_\ell\}\subset\hat{A}$, any cycle %\sknote{dans ce contexte, quand je parle de "cycle" je parle d'un ensemble d'arcs tels que, si on ne prend pas en compte leur orientation et qu'on les considère donc comme des arêtes, ces arêtes forment un cycle} in $(V,\hat{A}')$ must use $a_\ell$. 
Yet the existence of such a circuit is impossible, as it would entail that of a $u_\ell-v_\ell$ path $p'_\ell$ in $A'\backslash\{a_\ell\}\subset A$ with $p'_\ell\neq p_\ell$. Thus, $T'=(V,\hat{A}')$ is a spanning tree for $G$ and we let $\tilde{\mM}$ be the network matrix associated to $T'$ whose $i^{th}$ column $\tilde{\mathbf{c}_i}$ represents the unique $u_i-v_i$ path $p'_i$ in $T'$, for all $i$. We show that $\tilde{\mM}=\mM'$.\\
First, for all $i$ such that $a_i \in \hat{A}'$, we have $\mathbf{c}'_i=\tilde{\mathbf{c}}_i=\mathbf{e}_i.$
Now, let $a_i \in A\backslash\hat{A}'$. By hypothesis, $\mB=\mI_{n-1}+(\mathbf{c}_\ell-\mathbf{e}_k)\mathbf{e}_k^\top$ therefore $\mB^{-1}=\mI_{n-1}-\alpha_k(\mathbf{c}_\ell-\mathbf{e}_k)\mathbf{e}_k^\top$. Then, if $\mathbf{c}_i=\sum\limits_{j=1}^{n-1}\mu_j\mathbf{c}_j$ we find
$$
\mathbf{c}_i'=\mB^{-1}\mathbf{c}_i=\mathbf{c}_i-\alpha_k\mu_k(\mathbf{c}_\ell-\mathbf{e}_k)=\sum\limits_{j=1}^{n-1}(\mu_j-\alpha_k\mu_k\alpha_j)\mathbf{c}_j+\alpha_k\mu_k\mathbf{e}_k=\sum\limits_{\substack{j=1 \\ j \neq k}}^{n-1}(\mu_j-\alpha_k\mu_k\alpha_j)\mathbf{c}_j+\alpha_k\mu_k\mathbf{e}_k.
$$
Now, using the procedure given earlier to find the $u_i-v_i$ path $p_i' \in \hat{A}'$, we find that $\tilde{\mathbf{c}}_i$ must satisfy:
$$
\tilde{\mathbf{c}}_i=\sum\limits_{\substack{j=1 \\ j \neq k}}^{n-1}\mu_j\tilde{\mathbf{c}}_j+\mu_k\left(\sum\limits_{\substack{j\neq k \\ j=1}}^{n-1}\lambda_j\tilde{\mathbf{c}}_j+\lambda_\ell\tilde{\mathbf{c}}_\ell\right) \quad \text{where} \quad \mathbf{c}_k=\sum\limits_{\substack{j=1 \\ j \neq k}}^{n-1}\lambda_j\mathbf{c}_j.
$$
Using $\mathbf{c}_k=\sum\limits_{\substack{j=1 \\ j \neq k}}^{n-1}\lambda_j\mathbf{c}_j$ and $\mathbf{c}_\ell=\sum\limits_{j=1}^{n-1}\alpha_j$, we deduce that $\lambda_\ell=\alpha_k$ and that $\lambda_j=-\alpha_k\alpha_j$ when $j \neq k$. The expression above can then be simplified into
$$\tilde{\mathbf{c}}_i=\sum\limits_{\substack{j=1 \\ j\neq k}}^{n-1}(\mu_j+\mu_k\lambda_j)\tilde{\mathbf{c}}_j+\lambda_\ell\mu_k\tilde{\mathbf{c}}_\ell=\sum\limits_{\substack{j=1 \\ j\neq k}}^{n-1}(\mu_j-\mu_k\alpha_k\alpha_j)\tilde{\mathbf{c}}_j+\alpha_k\mu_k\tilde{\mathbf{c}}_\ell.$$
Since $\mathbf{c}'_j=\mathbf{e}_j=\tilde{\mathbf{c}}_j$ whenever $j\leq n-1,$ $j \neq k$ and since $\mathbf{c}'_\ell=\tilde{\mathbf{c}}_\ell=\mathbf{e}_k$, we deduce $\tilde{\mathbf{c}}_i=\mathbf{c}'_i$ for all $i$ and the result is proved.\\
Once again, although in the general case the matrix $\mB$ might not differ from $\mI_n$ by exactly one column, the reasoning can be applied iteratively. 

\fi
%
%\end{proof}

We now turn to the main result of this section, which establishes a direct 
connection between positive spanning sets and strongly connected digraphs. A 
similar result was stated without proof by Marcus~\cite{DAMarcus_1984}. 
However, to the best of our knowledge, Theorem~\ref{th:networkmatpss} in its 
present form and its proof are new.

\begin{theorem}
\label{th:networkmatpss} 
	Let $G=(V,A)$ be a connected digraph with $|A|=m$ and let
	$T=(V,\hat{A})$ be an oriented spanning tree of $G$ with $|\hat{A}|=n$. 
	Let $\mM \in \R^{n \times m}$ be the associated network matrix. 
	Then, the matrix $\mM$ is a PSS for $\R^n$ if and only if the graph 
	$G$ is strongly edge-connected.
\end{theorem}
\begin{proof}
	Without loss of generality, we consider an ordering $A=\{a_i\}_{i=1}^m$ 
	such that $\hat{A}=\{a_i\}_{i=1}^n$. As a result, the first $n$ columns of 
	$\mM$ form the identity matrix $\mI_n$, thus ensuring that $\mM$ has full 
	row rank. We now proceed with the proof.
	
	Suppose first that $G$ is strongly connected, and consider an ear 
	decomposition $\{G_i=(V_i,A_i)\}$ associated with $G$. Since 
	$G_1$ consists of a single vertex without arcs, there exists no circuit 
	within $G_1$. 
	Now, for any $i \in [\![2,k]\!]$, consider the ear of $G_i$ defined by $A_i\backslash A_{i-1}$. The arcs in this ear belong to a circuit of $G$ that 
	possibly contains other arcs in $G_{i-1}$. 
	Let $\vx_i \in \R^m$ be the characteristic vector of this circuit, i.e. 
	$[\vx_i]_j=1$ if arc $a_j$ is in the circuit, and $[\vx_i]_j=0$ otherwise. 
	\rev{Then the vector $\mM \vx_i$ corresponds to the sum of all arcs in this circuit, which by definition must be $\zero_n$.}
	Finally, consider the vector $\vx = \sum_{i=2}^s \vx_i$. This vector has 
	positive coefficients, since every arc in $G$ is contained in an ear of the decomposition.
	We have thus found a positive vector $\vx$ such that $\mM\vx=\zero_n$, hence 
	$\mM$ is a PSS by Proposition~\ref{prop:characterizepss}.
	
	Conversely, suppose that $G$ is \emph{not} strongly connected. We will show 
	that there exists a vector $\vy \in \R^n$ such that 
	$\vy^\T \mM \ge \zero_m$, thereby implying that $\mM$ is not a PSS thanks to 
	Proposition~\ref{prop:certifnopss}.
	Since $G$ is not strongly connected, 
	Proposition~\ref{prop:cutstrongconnected} ensures that there exists a set $\tilde{A} \subset A$ and two 
	subgraphs $G_1=(V_1,A_1)$, $G_2=(V_2,A_2)$ that define an oriented cut 
	of $G$. Then, since $G$ is connected, we find that $\tilde{A}\cap\hat{A}$ is non-empty. Without 
	loss of generality, we assume that $\tilde{A}\cap \hat{A} = \{a_1,\dots,a_k\}$ for 
	some $k \le n$.
	Consider the vector $\vy = \sum_{\ell=1}^k \vy_\ell$, where $\vy_\ell$ is the 
	$\ell^{th}$ column of $\mM$. Since the first $n$ columns of $\mM$ correspond to 
	the identity matrix, it follows that 
	\[
		\vy^\T \vy_i = 
		\left\{ 
		\begin{array}{ll}
			1 &\mbox{if $i \in [\![1,k]\!]$}\\
			0 &\mbox{if $i \in [\![k+1,n]\!]$}.
		\end{array}
		\right.
	\]
	In addition, if $i \in [\![n+1,m]\!]$, then two situations can occur. If 
	arc $a_i$ belongs to either 
	$A_1$ or $A_2$, then the directed path in $T$ linking the head and tail of $a_i$ 
	must thus contain an even number of arcs in $\tilde{A}$ (possibly $0$), with half of 
	them used in the forward direction. As a result, one must have 
	\[
		\vy^\T \vy_i = \sum_{\ell=1}^k \vy_{\ell}^\T \vy_i = 0.
	\]	
	Otherwise $a_i$ belongs to $A \setminus (A_1 \cup A_2)$, that is $a_i \in \tilde{A}$ therefore the tail 
	of $a_i$ is in $V_1$ and its head is in $V_2$. As a result, the directed path 
	in $T$ linking the head and tail of $a_i$ must use an odd number of arcs in 
	$\tilde{A} \cap \hat{A}$, with one more arc in the forward direction. Thus,
%	(i.e. the head of $a_i$ and the tail of $a_i$ belong to different subgraphs)
	\[
		\vy^\T \vy_i = \sum_{\ell=1}^k \vy_{\ell}^\T \vy_i = 1.
	\]	
	Overall, we have shown that $\vy^\T \mM \ge \zero_m$, from which we conclude 
	that $\mM$ cannot be a PSS.
\end{proof}

Theorem~\ref{th:networkmatpss} provides positive and negative certificates 
regarding the PSS (or non-PSS) nature of a network matrix based on those for 
strongly connected digraphs. We illustrate below the result of 
Theorem~\ref{th:networkmatpss} using two examples.

\begin{example}[A PSS network matrix]
\label{ex:PSSnetmatrix}
% Actually PB/minimally strongly connected
The positive spanning set $\mathbf{M_1}$ is a network matrix associated to the 
strongly connected digraph below and to the spanning tree formed by the thick arcs. 
\\[0.25cm]
\noindent % Prevents indentation
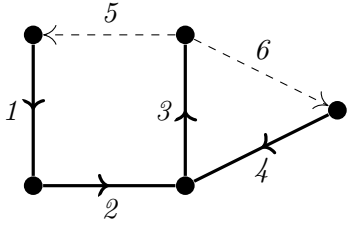
\begin{minipage}{0.5\textwidth} % Adjust width to your preference
  \begin{tikzpicture}
\tikzstyle{vertex}=[circle, fill=black, minimum size=7pt, inner sep=1pt]
\tikzstyle{vertex2}=[circle, fill=white, minimum size= 2pt, inner sep=1pt]
\tikzset{->-/.style={decoration={
  markings,
  mark=at position .5 with {\arrow{>}}},postaction={decorate}}}
\tikzstyle{edge}=[->-, very thick]
\node[vertex] (v1) at (0,0) {};
\node[vertex] (v2) at (2,0) {};
\node[vertex] (v3) at (4,1) {};
\node[vertex] (v4) at (2,2) {};
\node[vertex] (v5) at (0,2) {};
\draw[edge] (v1) to (v2);
\node[vertex2] (e2) at (1,-0.3) {2};
\draw[edge] (v2) to (v4);
\node[vertex2] (e3) at (1.7,1) {3};
\draw[edge] (v5) to (v1);
\node[vertex2] (e1) at (-0.3,1) {1};
\draw[edge] (v3) to (v2);
\node[vertex2] (e4) at (3,0.2) {4};
\draw[-{>[scale=2.5,
          length=2,
          width=3]},line width=0.4pt, dashed] (v4) to  (v3);
\node[vertex2] (e5) at (3,1.8) {6};
\draw[-{>[scale=2.5,
          length=2,
          width=3]},line width=0.4pt, dashed] (v4) to  (v5);
\node[vertex2] (e6) at (1,2.3) {5};
\end{tikzpicture} 
\end{minipage}
\hfill 
\begin{minipage}{0.5\textwidth}
  $\mathbf{M_1}=\bbm1&0&0&0&-1&0\\0&1&0&0&-1&0\\ 0&0&1&0&-1&-1 \\ 0&0&0&1&0&-1\ebm$
\end{minipage}\\[0.25cm]
\end{example}

\begin{example}[A non-PSS network matrix]
\label{ex:nonpssnetworkmat}
Let $T=(V,\hat{A})$ be the spanning tree formed by the thick arcs in the digraph $G$ below. The wavy cut $\tilde{A}$ certifies that $G$ is not strongly connected. Similarly, the network matrix $\mathbf{M_2}$ is not a PSS as the characteristic vector $\mathbf{y}^\top=\bbm1&1&0&0&0&0\ebm$ of $\tilde{A}\cap \hat{A}$ in $T$ satisfies $\mathbf{y}^\top\mathbf{M_2}\geq \zero_{10}.$
\\[0.25cm]

\noindent % Prevents indentation
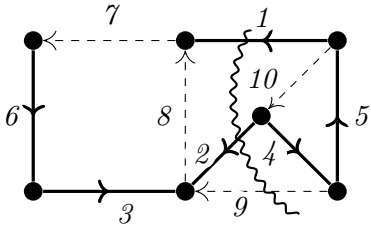
\begin{minipage}{0.5\textwidth} % Adjust width to your preference
  \begin{tikzpicture}
\tikzstyle{vertex}=[circle, fill=black, minimum size=7pt, inner sep=1pt]
\tikzstyle{vertex2}=[circle, fill=white, minimum size= 2pt, inner sep=1pt]
\tikzset{->-/.style={decoration={
  markings,
  mark=at position .5 with {\arrow{>}}},postaction={decorate}}}
\tikzstyle{edge}=[->-, very thick]
\node[vertex] (v1) at (0,0) {};
\node[vertex] (v2) at (2,0) {};
\node[vertex] (v3) at (3,1) {};
\node[vertex] (v4) at (4,0) {};
\node[vertex] (v5) at (4,2) {};
\node[vertex] (v6) at (0,2) {};
\node[vertex] (v7) at (2,2) {};
\draw[edge] (v6) to (v1);
\draw[edge] (v1) to (v2);
\draw[edge] (v3) to (v2);
\draw[-{>[scale=2.5,
          length=2,
          width=3]},line width=0.4pt, dashed] (v2) to  (v7);
\draw[edge] (v4) to (v5);
\draw[edge] (v3) to (v4);
\draw[edge] (v5) to (v7);
\draw[thick,decorate,decoration={snake,segment length=2mm,amplitude=0.4mm}] (3,2.4) to [bend right=45] (3.5,-0.3);
\draw[-{>[scale=2.5,
          length=2,
          width=3]},line width=0.4pt, dashed] (v4) to  (v2);
\draw[-{>[scale=2.5,
          length=2,
          width=3]},line width=0.4pt, dashed] (v7) to  (v6);
%\node[vertex2] (e5) at (3,1.8) {5};
\draw[-{>[scale=2.5,
          length=2,
          width=3]},line width=0.4pt, dashed] (v5) to  (v3);
%\node[vertex2] (e6) at (1,2.3) {6};

\node[vertex2] (e1) at (-0.3,1) {6};
\node[vertex2] (e2) at (1.2,-0.3) {3};
\node[vertex2] (e3) at (2.2,0.5) {2};
\node[vertex2] (e4) at (3.1,0.5) {4};
\node[vertex2] (e5) at (4.3,1) {5};
\node[vertex2] (e6) at (3,2.3) {1};
\node[vertex2] (e7) at (1,2.3) {7};
\node[vertex2] (e8) at (1.7,1) {8};
\node[vertex2] (e9) at (2.7,-0.2) {9};
\node[vertex2] (e10) at (3,1.5) {10};

\end{tikzpicture} 
\end{minipage}
\hfill 
\begin{minipage}{0.5\textwidth}
  $\mathbf{M_2}=
  \bbm
     1&0&0&0&0&0&-1& 1& 0& 0
  \\ 0&1&0&0&0&0& 1&-1& 1& 0
  \\ 0&0&1&0&0&0&-1& 0& 0& 0
  \\ 0&0&0&1&0&0&-1& 1&-1&-1
  \\ 0&0&0&0&1&0&-1& 1& 0&-1
  \\ 0&0&0&0&0&1&-1& 0& 0& 0
  
  \ebm$
\end{minipage}
\end{example}

In the next section, we will exploit the link between strongly connected 
digraphs and PSSs by showing that the ear decomposition can be extended to 
more general matrices.

%%%%%%%%%%%%%%%%%%%%%%%%%%%%%%%%%%%%%%%%%%%%%%%%%%%%%%%%%%%%%%%%%%%%%%%%%%%%%%%
\section{Ear decomposition of positive spanning sets}
\label{sec:eardec}
%%%%%%%%%%%%%%%%%%%%%%%%%%%%%%%%%%%%%%%%%%%%%%%%%%%%%%%%%%%%%%%%%%%%%%%%%%%%%%%

Theorem~\ref{th:eardecompograph} states that a digraph is strongly connected 
if and only if it admits an ear decomposition. Together with 
Theorem~\ref{th:networkmatpss}, it thus implies that a network matrix is a PSS 
if and only if its associated graph admits an ear decomposition. We will now 
establish a similar result for positive spanning sets, thereby generalizing 
Theorem~\ref{th:eardecompograph}. To this end, we start this section by 
extending the notions of circuits and acyclic graphs to matrices.

%%%%%%%%%%%%%%%%%%%%%%%%%%%%%%%%%%%%%%%%%%%%%%%%%%%%%%%%%%%%%%%%%%%%%%%%%%%%%%%
\subsection{Acyclic and circuit matrices}
\label{ssec:circacyclmat}

The ear decomposition for strongly connected digraphs relies on the 
fundamental notion of circuit. Our goal is thus to define an equivalent 
concept for matrices. To this end, we first introduce the companion notion of 
acyclic matrix, inspired by acyclic graphs.

\begin{definition}[Acyclic matrix]
\label{def:acyclic}
	A matrix $\mA \in \R^{n \times m}$ is called \emph{acyclic} if 
	\[
		\left[\ \mA \vx = \zero_n \ \mbox{and}\ \vx \ge \zero_m\ \right]
		\quad \Leftrightarrow \quad 
		\vx = \zero_m.
	\]
\end{definition}

Per Definition~\ref{def:acyclic}, any network matrix associated to an acyclic 
graph is an acyclic matrix. Alternate characterizations of acyclic matrices 
can be obtained from Gordan's Lemma~\cite{OMangasarian_1981}.

\begin{proposition}
\label{prop:acyclic}
	For any $\mA \in \R^{n \times m}$, the following statements are equivalent.
	\begin{enumerate}[(i)]
		\item\label{item1:propacyclic} The matrix $\mA$ is acyclic.
		\item\label{item2:propacyclic} There exists a non-zero vector $\vy \in \R^n$ 
		such that $\vy^\T \mA >\zero_m^\T.$
		\item\label{item3:propacyclic} For any $\bar{\mA} \in \R^{n \times \bar{m}}$ 
		with $\cols(\bar{\mA})\subset \cols(\mA)$ and $\bar{m}>0$, 
		$\pspan(\bar{\mA})$ is not a linear space.
	\end{enumerate}
\end{proposition}
\begin{proof}
	The equivalence between (i) and (ii) is a restatement of Gordan's lemma. The 
	equivalence between (i) and (iii) follows from that between 
	the first and last statements of Proposition~\ref{prop:characterizepss}.
\end{proof}

From Proposition~\ref{prop:acyclic}, one observes that matrices with positive 
entries are necessarily acyclic. In fact, positive entries characterize 
acyclic matrices in the following sense.

\begin{proposition}
\label{prop:posentriesacyclic}
	A matrix $\mA \in \R^{n\times m}$ is acyclic if and only if it is 
	structurally equivalent to a matrix with \rev{all entries strictly positive}.
\end{proposition}
\begin{proof}
	Suppose first that $\mA$ is acyclic. From Proposition~\ref{prop:acyclic}(ii), 
	there exists $\vy \in \R^n$ such that $\vy^\T \mA > \zero_m^\T$. Then, there 
	exists $\eta>0$ such that the matrix
	$\mB = \bbm \vy^\T\\ \dots \\ \vy^\T \ebm + \eta \mI_n \in \R^{n \times n}$ is 
	invertible and $\mB\mA$ has positive entries, proving the desired result. 
	
	Conversely, suppose that $\mA$ is not acyclic. Then, there must exist a 
	nonzero vector $\vx \ge \zero_m$ such that $\mA \vx=\zero_n$. For any 
	invertible matrix $\mB$, one then has $\mB^{-1}\mA\vx=\zero_n$, hence the 
	matrix $\mB^{-1}\mA$ must have \rev{at least} one non-positive entry.
\end{proof}

Proposition~\ref{prop:posentriesacyclic} provides a certificate for showing 
that a matrix is acyclic, by finding a basis defining a structurally 
equivalent matrix. A certificate for showing that a matrix is \emph{not} 
acyclic is obtained by finding a nonzero vector with nonnegative entries in the 
null space of the matrix.

We are now ready to provide the definition of a circuit matrix.

\begin{definition}[Circuit]
\label{de:circuitmat}
	A matrix $\mC \in \R^{n\times m}$ is called a \emph{circuit} matrix if it 
	is a PSS for some linear subspace $\LL$ of $\R^n$ and if any matrix 
	$\overline{\mC} \in \R^{n \times \bar{m}}$ formed by $0<\bar{m}<\dim(\LL)+1$ 
	columns of $\mC$ is acyclic.
\end{definition}

We will see in Section~\ref{sec:apps} that circuit matrices can be identified 
with a special class of PSSs called minimal positive bases. Those are 
instrumental in obtaining decompositions of PSSs, and we will use circuit 
matrices for a similar purpose in the next section. In particular, the 
following structural equivalence will be leveraged.

\begin{proposition}
\label{prop:circuitmatequiv}
	Let $\mC \in \R^{n \times m}$ be a circuit matrix. Then $\mC$ is a 
	PSS for a linear subspace of dimension $\ell=m-1$ therefore
	\begin{equation}
	\label{eq:circuitmatequiv}
		\mC 
		\; \equiv \;
		\bbm
			\mI_{\ell} &-\one_{\ell} \\
			\zero_{n-\ell,\ell} &\zero_{n-\ell}
		\ebm.
	\end{equation}
\end{proposition}

\begin{proof}
	By definition, $\mC$ is a PSS for some linear subspace $\LL$ of $\R^n$. 
	For the purpose of contradiction, suppose that $m>\ell+1$, where 
	$\ell=\dim(\LL)$. Since $\mC$ is a PSS for $\LL$, 
	Proposition~\ref{prop:characterizepss} ensures that there exists a positive 
	vector $\vx \in \R^n$ such that $\mC \vx = \zero_n$. In addition, there 
	also exists a matrix $\bar{\mC} \in \R^{n \times (m-1)}$ formed by 
	columns of $\mC$ such that $\spann(\bar{\mC})=\LL$. Without loss of generality, 
	suppose that $\bar{\mC}$ consists of the first $m-1$ columns of $\mC$. 
	Since $m-1>\ell$, the matrix $\bar{\mC}$ has a non-zero null space, i.e. 
	there exists a nonzero vector $\bar{\vy} \in \R^{m-1}$ such that 
	$\bar{\mC}\bar{\vy}=\zero_{m-1}$. Letting $\vy=[\bar{\vy}^\T\ 0]^\T$, it 
	follows that $\mC\vy=\zero_n$.
	
	Now, the vectors $\vx$ and $\vy$ are not colinear since 
	$[\vx]_m > 0 = [\vy]_m$. Therefore, there exists $\sigma \ge 0$ such that 
	$\vz = \sigma\vx+\vy \ge \zero_n$ with $\vz$ having at least one component equal 
	to zero. Let $\tilde{\mC}$ be the matrix formed by columns of $\mC$ 
	corresponding to non-zero components of $\vz$, and let $\tilde{\vz}$ be 
	the vector formed by those components. It follows from the construction 
	of $(\tilde{\mC},\tilde{\vz})$ that
	\[
		\tilde{\mC}\tilde{\vz}=\mC\vz=\mC(\sigma\vx+\vy)=\zero_n.
	\]
	As a result, we have shown that $\tilde{\mC}$ satisfies 
	Proposition~\ref{prop:characterizepss}, and thus it must be a PSS for 
	some linear space of $\R^n$. Therefore, $\tilde{\mC}$ is not acyclic, and 
	this contradicts the fact that $\mC$ is a circuit, from which we conclude 
	that $m=\ell+1$.

	The second part of the result then follows from the special 
	case~\eqref{eq:nicestructureminpb} in Proposition~\ref{prop:nicestructurepss}.
\end{proof}	
	
Note that Proposition~\ref{prop:circuitmatequiv} justifies the terminology 
\emph{circuit matrix}, as a circuit of $n+1$ vertices admits 
$\bbm \mI_n&-\one_n\ebm$ as a network matrix. Using structural equivalence for
circuits, we can obtain an improved certificate for non-acyclic matrices.

\begin{lemma}
\label{le:certifnoacycl}
	A matrix $\mM \in \R^{n \times m}$ is not acyclic if and only if 
	\begin{enumerate}[(i)]
		\item One of the columns of $\mM$ is the zero vector $\zero_n$, or
		\item There exists $\ell \in [\![1,n]\!]$ such that
		\[
			\mM \equiv 
			\bbm 
			\mI_{\ell}			&-\one_\ell		&\mX \\ 
			\zero_{n-\ell,\ell}	&\zero_{n-\ell}	&\mY
			\ebm,
		\] 
		where the matrices $\mX \in \R^{\ell \times (m-\ell-1)}$ and 
		$\mY\in \R^{(n-\ell)\times (m-\ell-1)}$ can be empty.
	\end{enumerate}
\end{lemma}
\begin{proof}
	If (i) holds, then the matrix $\mM$ is not acyclic as 
	Proposition~\ref{prop:acyclic}(ii) fails. If (ii) holds, then the 
	first $\ell+1$ columns of $\mM$ form a matrix $\bar{\mM}$ such 
	that $\pspan(\bar{\mM})$ is a linear space. Thus, 
	Proposition~\ref{prop:acyclic}(iii) fails to hold, hence $\mM$ 
	is not acyclic.
	Conversely, suppose that $\mM$ is not acyclic. Then by 
	Proposition~\ref{prop:acyclic}(iii), there must exist a matrix 
	$\mC$ formed by a subset of columns of $\mM$ such that 
	$\pspan(\mC)$ is a linear space. Without loss of 
	generality, suppose that this matrix is minimal for that 
	property, and that it consists of the first $\ell+1$ columns of 
	$\mM$ (thus the linear subspace spanned by $\mC$ is of 
	dimension $\ell$). Then the matrix $\mC$ is a circuit. Applying 
	Proposition~\ref{prop:circuitmatequiv}, we then know that there exists 
	an invertible matrix $\mB$, a permutation matrix 
	$\mP$ and a diagonal matrix $\mathbf{\Delta}$ with positive diagonal entries such that
	\[
		\mB^{-1}\mC\mP\mathbf{\Delta}=
		\bbm 
		\mI_{\ell} 				&-\one_\ell \\ 
		\zero_{n-\ell,\ell}^\T	&\zero_{n-\ell}
		\ebm.
	\]
	Letting then $\mP' = \bbm\mathbf{P\Delta}&\zero_{\ell+1,m-\ell-1}\\ 
	\zero_{m-\ell-1,\ell+1}&\mathbf{I_{m-\ell-1}}\ebm \in \R^{m \times m}$, 
	it follows that $\mM \equiv \mB^{-1}\mM \mP'$, that has the desired 
	structure.
\end{proof}

A consequence of Lemma~\ref{le:certifnoacycl} is that the vector 
$\vx=\bbm\one_{\ell+1}\\ \zero_{m-\ell-1}\ebm$ can be used to attest that 
$\mM$ is not acyclic by structural equivalence, thus improving over the 
certificate from Proposition~\ref{prop:posentriesacyclic}.

%%%%%%%%%%%%%%%%%%%%%%%%%%%%%%%%%%%%%%%%%%%%%%%%%%%%%%%%%%%%%%%%%%%%%%%%%%%%%%%
\subsection{A new certificate for positive spanning sets}
\label{ssec:certifPSS}

Building on the results of the previous subsection, we now generalize the 
concept of ear decomposition of digraphs to matrices. As a result, we will 
obtain a characterization of PSSs improving that of 
Proposition~\ref{prop:nicestructurepss}.

\begin{definition}[Negative row echelon matrix]
\label{def:negativeechelonform}
	A matrix $\mN \in \R^{n \times s}$, with $1 \le s \le n$, is a 
	\emph{negative row echelon matrix} (NEM) if there exists a 
	sequence $z_0 = 1 < z_1 < z_2 < \dots < z_{s-1} \leq n$ such that
	\begin{enumerate}[(i)]
		\item For all $j \in [\![1,s-1]\!]$, 
		for all $i \in [\![z_j,n]\!]$, 
		$\mN_{i,j}=0$.
		\item For all $j \in [\![1,s-1]\!]$, 
		for all $i \in [\![z_{j-1},z_j-1]\!]$, 
		$\mN_{i,j}=-1$.
		\item $\mN_{i,s} = - 1$, for all $i\geq z_{s-1}$.
	\end{enumerate}
\end{definition}

An example of NEM is 
\[
	\bbm 
	-1 & \times & \times & \times  \\
	-1 & \times & \times & \times  \\
	0 &    -1   & \times & \times  \\
	0 &    -1   & \times & \times  \\
	0 &    0   &    -1   & \times \\
	0 &    0   &    0   &    -1
	\ebm,
\] 
where the crosses $\times$ indicate arbitrary values. Matrices of that form can be 
used to create PSSs as follows.

\begin{proposition}
\label{prop:PSSNEM}
	Let $\mM \in \R^{n\times (n+s)}$ with $s \in [\![1,n]\!]$ 
	such that $\mM \equiv \bbm \mI_n &\mN \ebm$ 
	where $\mN$ is a NEM. Then, $\mM$ is a PSS.
\end{proposition}
\begin{proof}
	Let $\mN = \bbm \vu_1 &\cdots &\vu_s \ebm$. 
	Consider the vectors $\{\vw_i\}_{i=1,\dots,s}$ defined by 
	\begin{eqnarray*}
		\vw_1 &= &\vu_s \\
		\forall i=1,\dots,s-1, \quad 
		\vw_{i+1} &= &\vw_i + 2 \|\vw_i\|_{\infty}\vu_{s-i}.
	\end{eqnarray*}	
	By construction, the vector $\vw_s$ is a positive linear combination of the 
	columns of $\mN$ and all its components are negative. 
	It follows that there exists a positive combination of columns of 
	$\bbm \mI_n\ \mN \ebm$ adding to $\zero_n$, i.e. a vector $\vx \in \R^{n+s}$ with positive 
	coefficients such that $\bbm \mI_n\ \mN \ebm\vx=\zero_n$. By 
	Proposition~\ref{prop:characterizepss}(iii) and structural equivalence, this 
	implies that $\mM$ is also a PSS.
\end{proof}

Proposition~\ref{prop:PSSNEM} complements 
Proposition~\ref{prop:nicestructurepss} in that it provides a sufficient 
condition for a matrix to be a PSS. Note that the result can be generalized to 
any matrix $\mM$ such that 
$
	\mM 
	\equiv 
	\bbm 
	\mI_{\ell} & \mathbf{N} \\ 
	\zero_{n-\ell,\ell}&\zero_{n-\ell,m-\ell}
	\ebm
$ 
where $\mN$ is NEM. Such a matrix $\mM$ is indeed a PSS for some 
$\ell$-dimensional linear space.

We now aim at providing certificates for determining whether a matrix is a PSS 
using NEMs, based on identifying a desirable structure within the matrix. 
Recall from Gaussian elimination that any matrix in $\R^{n \times m}$ is 
structurally equivalent to a matrix of the form
\[
	\bbm
		\mI_{\ell} &\mX \\
		\zero_{n-\ell,\ell} &\zero_{n-\ell,m-\ell}
	\ebm,
\]
for some $\ell \in [\![0,n]\!]$ and some matrix 
$\mX \in \R^{\ell \times (m-\ell)}$. When $\ell=n$ and $\mX$ is a NEM, such 
a structure allows to conclude that the matrix is a spanning set of $\R^n$. 
Following Proposition~\ref{prop:PSSNEM}, we now define structures of interest 
for positive spanning sets.

\begin{definition}%[INX matrix]
\label{de:ININA}
	Given dimensions $n,m$, let $\ell \in [\![1,n]\!]$ and 
	$k \in [\![0,m-\ell]\!]$.
	\begin{enumerate}[(i)]
		\item An \emph{IN matrix} in $\R^{n \times m}$ is a matrix of the form
		\begin{equation}
		\label{eq:INmat}
			\bbm
				\mI_{n,\ell} &\mN &\mX 
			\ebm,
		\end{equation}
		where $\mI_{n,\ell}$ represents the first $\ell$ columns of the identity 
		matrix in $\R^{n \times n}$, $k \le n$, $\mN \in \R^{n \times k}$ is a 
		NEM when $k>0$ and $\mX \in \R^{n \times (m-\ell-k)}$ is arbitrary.
		\item An \emph{INA matrix} in $\R^{n \times m}$ is a matrix of the form
		\begin{equation}
		\label{eq:INAmat}
			\bbm
				\mI_{\ell} &\mN &\mX \\
				\zero_{n-\ell,\ell} &\zero_{n-\ell,k} &\mA
			\ebm,
		\end{equation}
		where $k \le \ell<n$, $\mN \in \R^{\ell \times k}$ is a NEM when $k>0$, 
		$\mX \in \R^{n \times (m-\ell-k)}$ is arbitrary and 
		$\mA \in \R^{(n-\ell) \times (m-\ell-k)}$ is an acyclic matrix.
	\end{enumerate}
\end{definition}

\rev{The matrix $\mM_1$ from Example~\ref{ex:PSSnetmatrix} is an example of an IN 
matrix, while the network matrix from Figure~\ref{fig:digraph and netwmat} is an 
INA matrix. Moreover,} since the definition of IN and INA matrices allows for $k=0$, 
any nonzero matrix is structurally equivalent to an IN matrix with $k=0$, per the 
Gaussian elimination argument above. However, the NEM and acyclic components in IN and 
INA matrices, respectively, allow for identifying positive spanning properties 
(or lack thereof) of a matrix. This is the purpose of the following theorem, 
that forms the central result of our paper.

\begin{theorem}
\label{th:structPSSIN}
	Let $\mM \in \R^{n \times m}$ be a nonzero matrix with $n \ge 2$.
	\begin{enumerate}[(i)]
		\item $\mM$ is a PSS of $\R^n$ if and only if $\mM$ is structurally 
		equivalent to an IN matrix with $\ell=n$ and $k >0$.
		\item $\mM$ is not a PSS of $\R^n$ if and only $\mM$ is structurally 
		equivalent to an INA matrix.
	\end{enumerate}
\end{theorem}

\begin{proof}
For both results, the reverse implication is immediate. Indeed, if $\mM$ 
is structurally equivalent to an IN matrix with $\ell=n$ and $k>0$, 
Proposition~\ref{prop:PSSNEM} guarantees that this IN matrix contains a 
PSS of $\R^n$, thus both this IN matrix and $\mM$ are PSSs of $\R^n$. 
In addition, if $\mM$ is structurally equivalent to an INA matrix, Proposition~\ref{prop:posentriesacyclic} implies that $\mM$ is 
also structurally equivalent to a matrix whose last $n-\ell$ rows are 
nonnegative.

In the rest of the proof, we thus focus on establishing the forward 
implication for both (i) and (ii) through an induction argument on 
the dimension $m$. Note that $m<n+1$ implies that $\mM$ cannot be a 
PSS and that it is structurally equivalent to an INA matrix with $k=0$.

% n=2
Suppose first that $m=n+1$. On one hand, if $\mM$ is a PSS, then it is also a 
circuit matrix with $\ell=n$. Proposition~\ref{prop:circuitmatequiv} then 
ensures that $\mM$ is equivalent to an IN matrix with $\ell=n$ and $k=1$. 
On the other hand, if $\mM$ is not a PSS it is either acyclic - and structurally 
equivalent to an INA matrix with $\ell=k=0$ - or it contains a circuit, say of 
length $\ell$. We choose this circuit such that $\ell<n$ is as large as possible.
From Proposition~\ref{prop:circuitmatequiv}, it follows that 
$\mM \equiv 
\bbm \mI_\ell &-\one_\ell&\bar{\mX} \\ 
\zero_{n-\ell,\ell} &\zero_{n-\ell}&\bar{\mA} 
\ebm$, where $\bar{\mX} \in \R^{\ell \times (n-\ell)}$ is arbitrary and 
$\bar{\mA} \in \R^{(n-\ell) \times (n-\ell)}$ is acyclic from the definition 
of $\ell$. It follows that $\mM$ is equivalent to an INA matrix with $k=1$, and 
we have thus shown that (i) and (ii) hold for $m=n+1$ whenever $n\geq2$.

Suppose now that $m>n+1$, and that (i) and (ii) hold for any $n \ge 2$ and 
any $\tilde{m} \in [\![n+1,m-1]\!]$. We will establish that (i) and (ii) 
hold as well.

\paragraph{On the one hand, suppose that $\mM$ is a PSS of $\R^n$.} 
If there exists a strict subset of columns of $\mM$ that is a PSS of $\R^n$, 
let $\bar{\mM} \in \R^{n \times \bar{m}}$ be a matrix formed by these columns. 
Then, $\mM \equiv \bbm \bar{\mM} &\mX_1 \ebm$ for some matrix 
$\mX_1 \in \R^{n \times (m-\bar{m})}$. Since $\bar{m}<m$ by assumption, 
the induction argument applied to $\bar{\mM}$ guarantees that $\bar{\mM}$ 
is equivalent to an IN matrix, i.e.
\[
	\bar{\mM} 
	\equiv 
	\bbm 
	\mI_n &\bar{\mN} &\bar{\mX} 
	\ebm,
\]
where $\bar{\mN} \in \R^{n \times k}$ is a NEM with $k>0$. Letting 
$\mX = \bbm \bar{\mX} &\mX_1 \ebm$, it follows that $\mM$ is 
equivalent to the IN matrix $\bbm \mI_n &\bar{\mN} &\mX \ebm$.\\ 
If no strict subset of columns of $\mM$ is a PSS of $\R^n$, we use 
the fact that $\mM$ is a PSS, thus it is not acyclic.
Lemma~\ref{le:certifnoacycl} then guarantees that there exists 
$\ell_1 \in [\![1,n]\!]$ such that 
\begin{equation}
\label{eq:decompMbarM}
	\mM 
	\equiv 
	\bbm 
	\mI_{\ell_1} &-\one_{\ell_1} &\mX_1 \\
	\zero_{n-\ell_1,\ell_1} &\zero_{n-\ell_1} &\bar{\mM} 
	\ebm.
\end{equation}
where $\bar{\mM} \in \R^{(n-\ell_1) \times \bar{m}}$  with 
$\bar{m}:=m-(\ell_1+1) \in [\![(n-\ell_1)+1,m-1]\!]$. 
Since $\mM$ is a PSS of $\R^n$, it follows 
from~\eqref{eq:decompMbarM} that $\bar{\mM}$ must be a PSS of 
$\R^{n-\ell_1}$. Applying the induction argument to 
$\bar{\mM}$ yields the structural equivalence
\[
	\bar{\mM} 
	\equiv
	\bbm
	\mI_{n-\ell_1} &\bar{\mN} &\bar{\mX} 
	\ebm,
\]
where $\bar{\mN}$ is a NEM, and $\bar{\mX}$ is arbitrary. With an 
appropriate change of basis, the following equivalence holds:
\[
	\bbm 
	\mI_{\ell_1} &\mX_1 \\
	\zero_{n-\ell_1,\ell_1} &\bar{\mM} 
	\ebm
	\equiv
	\bbm
	\mI_{\ell_1} &\zero_{\ell_1,n-\ell_1} &\mX_2 &\mX_3 \\
	\zero_{n-\ell_1,\ell_1} &\mI_{n-\ell_1} &\mN_1 &\mX_4,
	\ebm
\]
where $\mN_1$ is a NEM and $\mX_2$, $\mX_3$, $\mX_4$ are 
arbitrary. It follows that
\[
	\mM 
	\equiv 
	\bbm 
	\mI_n &\mN &\mX 
	\ebm
	\qquad
	\mbox{where} 
	\quad
	\mN 
	= 
	\bbm
		-\one_{\ell_1} &\mX_2 \\
		\zero_{n-\ell_1} &\mN_1
	\ebm
	\quad
	\mbox{and}
	\quad
	\mX
	=
	\bbm
		\mX_3 \\
		\mX_4
	\ebm.
\]
Since $\mN$ is a NEM by construction, we have shown that $\mM$ is equivalent to an IN matrix with $\ell=n$ and $k>0$.

\paragraph{On the other hand, suppose that $\mM$ is a not a PSS of $\R^n$\rev{.}} 
If $\mM$ is acyclic, then it is an INA matrix 
with $\ell=k=0$. 
Otherwise, Lemma~\ref{le:certifnoacycl} guarantees that there exists $\ell_1 \in [\![1,n]\!]$ such that $\mM \equiv \bbm \mI_{\ell_1} &-\one_{\ell_1}&\mM_1 \\ \zero_{n-\ell_1,\ell_1} &\zero_{n-\ell_1}&\mM_2 \ebm$, where $\mM_2 \in \R^{n-\ell_1 \times (m-\ell_1)}$ does not positively span $\R^{n-\ell}$ as $\mM$ is not a PSS of $\R^n$. Applying the induction argument, it follows that $\mM_2$ is structurally equivalent to an INA matrix, i.e. $$\mM_2\equiv\bbm\mI_{\bar{\ell}} &\bar{\mN} &\bar{\mX} \\\zero_{\tilde{\ell},\bar{\ell}} &\zero_{\tilde{\ell}} &\mA\ebm\quad \text{and so}\quad \mM\equiv \bbm \mI_{\ell_1}&-\one_{\ell_1}&\zero_{\ell_1,\bar{\ell}}&\mX_1&\mX_2\\ \zero_{\bar{\ell},\ell_1}&\zero_{\bar{\ell}}&\mI_{\bar{\ell}}&\bar{\mN}&\bar{\mX}\\\zero_{\tilde{\ell},\ell_1}&\zero_{\tilde{\ell}}&\zero_{\tilde{\ell},\bar{\ell}}&\zero_{\tilde{\ell},k}&\mA\ebm$$ with  $\bar{\ell}<n-\ell_1$, $\tilde{\ell}=n-\ell_1-\bar{\ell}$, $\bar{\mN}\in\R^{\bar{\ell}\times k}$ a NEM and $\mA$ an acyclic matrix.

Letting $\mX = \bbm \bar{\mX}_2\\\bar{\mX}\ebm$ and 
$\mN=\bbm -\one_{\ell_1}&\mX_1\\\zero_{\bar{\ell}}&\bar{\mN}\ebm$, we obtain that 
$\mM \equiv
\bbm\mI_{n-\tilde{\ell}} &\mN &\mX \\
\zero_{\tilde{\ell},\bar{\ell}} &\zero_{\tilde{\ell}} &\mA
\ebm,$ 
hence $\mM$ is equivalent to an INA matrix with $\ell=\bar{\ell}<n$.
\end{proof}

\begin{remark}
	Although Theorem~\ref{th:structPSSIN} assumes $n \ge 2$, a similar result 
	can be stated when $n=1$. Indeed, given any nonzero matrix 
	$\mM \in \R^{1 \times m}$, is it clear that $\mM$ is a PSS of $\R$ if and 
	only if $\mM \equiv \bbm 1 &-1 &\mX \ebm$ for some arbitrary 
	$\mX \in \R^{1 \times (m-2)}$, while it is not a PSS if and only if 
	$\mM \equiv \bbm 1 &\mA \ebm$ for some matrix 
	$\mA \in \R^{1 \times (m-1)}$ with nonnegative coefficients.
\end{remark}

Akin to the proof of Theorem~\ref{th:eardecompograph}, the proof of 
Theorem~\ref{th:structPSSIN} relies on an induction argument based on any 
column of a PSS being part of a circuit. The latter proof can thus be 
viewed as a generalization of the former. In the next section, we will study 
implications of Theorem~\ref{th:structPSSIN} on the characterization of 
certain PSSs called positive bases.

%%%%%%%%%%%%%%%%%%%%%%%%%%%%%%%%%%%%%%%%%%%%%%%%%%%%%%%%%%%%%%%%%%%%%%%%%%%%%%%
\section{Applications to positive bases}
\label{sec:apps}
%%%%%%%%%%%%%%%%%%%%%%%%%%%%%%%%%%%%%%%%%%%%%%%%%%%%%%%%%%%%%%%%%%%%%%%%%%%%%%%

Positive bases can be succinctly defined as inclusion-wise minimal positive 
spanning sets~\cite[Chapter 2]{ARConn_KScheinberg_LNVicente_2009b}. We begin 
this section by reviewing the concept of positive bases, drawing connections 
with digraphs as in Section~\ref{ssec:network}. We then provide a general 
characterization of positive bases.

%%%%%%%%%%%%%%%%%%%%%%%%%%%%%%%%%%%%%%%%%%%%%%%%%%%%%%%%%%%%%%%%%%%%%%%%%%%%%%%
\subsection{Positive bases and digraphs}
\label{ssec:posbases}

In this paper, we follow the notations of Hare et 
\rev{al.}~\cite{WHare_GJarryBolduc_SKerleau_CWRoyer_2024}, and define a positive 
basis by explicitly highlighting its associated subspace and size.

\begin{definition}[Positive basis] \label{def:pb}
	Let $\LL$ be an $\ell$-dimensional linear subspace of $\R^n$ with 
	$\ell \ge 1$. A matrix $\mD \in \R^{n \times (\ell+s)}$ 
    \rev{with $s\in[\![1,\ell]\!]$} is called a 
	\emph{positive basis} of $\LL$ of size $\ell+s$ if it is a PSS of $\LL$
	such that no proper subset of the columns of $\mD$ is a PSS for $\LL$. 

    We let $\mD_{\LL,s}$ denote such a positive basis. When $\LL=\R^n$, 
    we use the simplified notation $\mD_{n,s}$.
\end{definition}

\rev{Definition~\ref{def:pb} exploits the well-known fact that positive bases 
of $\LL$ have cardinality in $[\![\dim(\LL)+1,2\,\dim(\LL)]\!]$~\cite{CAudet_WHare_2017,CDavis_1954}.}
%The restriction on the values of the integer $s$ comes from the fact that positive bases of larger sizes cannot exist~\cite{CAudet_WHare_2017,CDavis_1954}. 
We say that \rev{a} positive basis is \emph{maximal} if $s=\ell$, 
\emph{minimal} when $s=1$, and \emph{intermediate} otherwise. The 
structure of the former two categories is well 
understood~\cite{CAudet_WHare_2017,RGRegis_2016}, and can be stated using 
structural equivalence as follows.

\begin{theorem}\label{th:structminmaxpb}
	Let $\LL$ be an $\ell$-dimensional linear subspace of $\R^n$, and let 
	$\mD_{\LL,1}$ and $\mD_{\LL,\ell}$ be a minimal positive basis and a 
	maximal positive basis of $\LL$, respectively. Then,
	\begin{equation*}
	\label{eq:structminmaxpb}
		\mD_{\LL,1} \equiv 
		\bbm 
			\mI_\ell& -\one_\ell \\ 
			\zero_{n-\ell,\ell}&\zero_{n-\ell}
		\ebm 
		\quad \mbox{and} \quad
		\mD_{\LL,\ell} \equiv 
		\bbm 
			\mI_\ell& -\mI_\ell \\ 
			\zero_{n-\ell,\ell}&\zero_{n-\ell,\ell}
		\ebm
	\end{equation*}
\end{theorem}

Theorem~\ref{th:structminmaxpb} shows that all minimal and maximal positive 
bases are structurally equivalent to simple matrices described through 
coordinate vectors and negative combinations thereof. In the case of minimal 
positive bases, Theorem~\ref{th:structminmaxpb} is actually a restatement of 
Proposition~\ref{prop:nicestructurepss}. 

Recall from Section~\ref{ssec:network} that the network matrix of a 
strongly connected digraph is a PSS. In particular, as stated in 
Section~\ref{ssec:circacyclmat}, a circuit $G$ on $n+1$ vertices 
admits the minimal positive basis $\bbm \mathbf{I_n}&-\one_n\ebm$ as a 
network matrix. Similarly, the maximal positive basis 
$\bbm \mathbf{I_n}&-\mathbf{I_n}\ebm$ is associated to a bi-directed tree 
on $n+1$ vertices. More generally, Theorem~\ref{th:networkmatpss} yields 
the following relationship between positive bases and minimally strongly 
connected digraphs.

\begin{corollary}~\label{cor:networkmatpbasis}
    Let $G=(V,A)$ be a connected digraph with $|V|=n$ and let
	$T=(V,\hat{A})$ be an oriented spanning tree of $G$. 
	Let $\mD \in \R^{(n-1) \times (n-1+s)}$ be a network matrix associated with 
	$\{G,T\}$. 
	Then, $\mD$ is a positive basis for $\R^{n-1}$ if and only if
	$G$ is minimally strongly connected.
\end{corollary}

\begin{remark}
    Using the bound on the size of positive bases, we note that
    Corollary~\ref{cor:networkmatpbasis} can be used to prove that the 
    number of arcs in a minimally strongly edge-connected digraph on $n$ 
    vertices ranges from $n$ to $2(n-1)$~\cite{JGarciaLopez_CMarijuan_2012}. 
    To the best of our knowledge, such a proof technique is novel.
\end{remark}
Given the link between positive bases and minimally strongly 
connected digraphs, we seek a characterization of positive bases. 
Adapting Theorem~\ref{th:structPSSIN} to such matrices, one sees that 
any positive basis $\mD_{n,s}$ satisfies the \rev{structural} equivalence
\begin{equation}
\label{eq:pbIN}
    \mD_{n,s} 
    \equiv 
    \begin{bmatrix} 
        \mI_n &\mN 
    \end{bmatrix}
    \quad \text{where} \quad
    \mN \in \R^{n \times s} \mbox{ is a NEM}.
\end{equation}
This structural characterization leads to the following result.

\begin{lemma}~\label{lem:nicenetworkmatrix}
Let $G=(V,A)$ be minimally strongly connected, let $T=(V,\hat{A})$ be a spanning 
tree of $G$, and let $\mD_{n,s} \in \R^{n\times (n+s)}$ be the network matrix 
associated with $(G,T)$. Then, for any matrix $\bbm\mI_n&\mN\ebm$ given 
by~\eqref{eq:pbIN}, the matrix $sgn(\bbm\mI_n&\mN\ebm)$ is also a network matrix 
associated with $G$ and a spanning tree $T'=(V,\hat{A}')$ with $\hat{A}'\subset A$.
\end{lemma}
\begin{proof}
By structural equivalence, the matrix $\mM=\bbm\mI_n&\mN\ebm$ can be written
$\mM=\mB^{-1}\mD_{n,s}\mP\Delta$ where $\mB \in \R^{n\times n}$ is nonsingular, 
$\mP \in \R^{(n+s)\times (n+s)}$ is a permutation matrix and 
$\Delta=\diag(\delta_1,\dots,\delta_{n+s})$ is a diagonal matrix with positive 
diagonal entries. Letting 
$\tilde{\Delta}=\diag(\delta_1,\dots,\delta_n) \in \R^{n\times n}$, we find that 
$\tilde{\Delta}\mB^{-1}\mD_{n,s}\mP=\bbm\mI_n&\tilde{\mN}\ebm$, where  
$sgn(\tilde{\mN})=sgn(\mN)$. It then follows from 
Proposition~\ref{prop: network matrices all equiv} that 
$\tilde{\Delta}\mB^{-1}\mD_{n,s}\mP$ is a network matrix and thus 
$\tilde{\Delta}\mB^{-1}\mD_{n,s}\mP 
= sgn(\tilde{\Delta}\mB^{-1}\mD_{n,s}\mP) 
= sgn(\mM)$.
\end{proof}

Unlike Theorem~\ref{th:structPSSIN} however, property \eqref{eq:pbIN} does not 
provide a full characterization of positive bases. For instance, the PSS 
$\begin{bmatrix} \mI_2 &-\ve_1 &-\one_2 \end{bmatrix}$ 
satisfies~\eqref{eq:pbIN} but is not minimal (the third vector can be 
removed without losing the positive spanning property).

Deriving a characterization thus requires a more precise study of the IN matrices 
that correspond to positive bases, which is the purpose of the next sections. 

%%%%%%%%%%%%%%%%%%%%%%%%%%%%%%%%%%%%%%%%%%%%%%%%%%%%%%%%%%%%%%%%%%%%%%%%%%%%%%%
\subsection{\rev{Critical structures for positive spanning sets}}
\label{ssec:crit}

A characterization of positive bases was given by Zbigniew 
Romanowicz based on the concept of critical vectors~\cite{ZRomanowicz_1987}. 
Partly due to this concept, this decomposition has proven difficult to use 
for characterizing positive bases. We redefine the notion of criticality 
below, and show that it simplifies in the case of NEMs and IN matrices. To 
this end, we will make use of special cones in $\R^n$. For any 
$i \in [\![1,n]\!]$, the cone
$
    \calK_i(\R^n) := \pspan\left(\bigcup\limits_{k \neq i}-\mathbf{e_k}\right) 
$
consist of vectors with non-positive coordinates with at least \rev{the 
$i^{\mathrm{th}}$ coordinate being} equal to zero. For any pair 
$(i,j) \in [\![1,n]\!]$ with $i<j$, the cone 
$
    \calK_{i,j}(\R^n) := \pspan\left(
    \{\one_n\}\cup 
    \bigcup\limits_{k \notin\{i,j\}}\{-\mathbf{e_k}\}
    \right)
$
consists of vectors \rev{whose $i^{\mathrm{th}}$ and $j^{\mathrm{th}}$ 
coordinates are both non-negative and equal to the larges entry of the 
vector}. Considering 
all cones above leads to the definition of critical vectors.

\begin{definition}[Critical vectors and critical matrix]
\label{de:crit}
The set of \emph{critical vectors} in $\R^n$, denoted by $\calK(\R^n)$, is defined 
as the union of all $\{\calK_i(\R^n)\}$ and $\{\calK_{i,j}(\R^n)\}$. 
A matrix $\mX \in \R^{n \times m}$ is called \emph{critical} whenever 
$\pspan(\mX) \subset \mathcal{K}(\R^n)$.
\end{definition}

We emphasize that the set $\calK(\R^n)$ is not a cone, which partly explains that 
the notion of critical vector is difficult to manipulate.

We now describe the link between our definition of critical vectors and that of 
Romanowicz~\cite{ZRomanowicz_1987}. In the latter, a vector $\vx \in \R^n$ is 
called critical \emph{with respect to} a positive basis $\mD_{n,s}$ whenever no positive 
spanning set can be obtained by substituting a column of $\mD_{n,s}$ with the 
vector $\vx$. As explained in the proposition below, those definitions are 
equivalent.
\begin{proposition}~\label{prop: crit vector D_n}
A vector $\vv \in \R^n$ is critical in the sense of 
Definition~\ref{de:crit} if and only if for all $k \in [\![1,n+1]\!]$, 
the matrix obtained by replacing the $k^{th}$ column of 
$\bbm\mathbf{I_n}&-\one_n\ebm$ with $\mathbf{v}$ is \emph{not} positively 
spanning.
\end{proposition}
\begin{proof}
Let $\vv \in \R^n$, $\mM_{\rev{n+1}}=\bbm \mI_n&\vv\ebm$. For any 
$k \in [\![1,n]\!]$, let $\mM_k$ be the matrix obtained from 
$\bbm \mI_n&-\one_n\ebm$ by replacing its $k^{th}$ column with $\vv$.

Suppose that $\vv$ is a critical vector. On one hand, if 
$\vv \in \mathcal{K}_i(\R^n)$ for some $i \in [\![1,n]\!]$, 
Proposition~\ref{prop:certifnopss} entails that $\mM_{\rev{n+1}}$ and $\mM_k$ are not 
positively spanning, as certified by the vectors $\mathbf{e_i}$ 
and $-\mathbf{e_k}$, respectively. On the other hand, if there exist 
$(i,j) \in [\![1,n]\!]^2$ with $i<j$ such that 
$\vv \in \calK_{i,j}(\R^n)\setminus \calK_i(\R^n)$, then $\vv$ satisfies 
$[\mathbf{v}]_i=[\mathbf{v}]_j
=\max\limits_{\ell \in [\![1,n]\!]}[\mathbf{v}]_\ell> 0$. Without loss of 
generality, suppose that $i\neq k$. Then, Proposition~\ref{prop:certifnopss} 
certifies that $\mM_{\rev{n+1}}$ and $\mM_k$ are not positively spanning, 
using the vectors $\mathbf{e_i}$ and $\mathbf{e_i}-\mathbf{e_k}$, respectively.

Conversely, suppose that $\vv$ is not a critical vector. If all coordinates of 
$\vv$ are negative, then $\mM_{\rev{n+1}}$ is a PSS. Otherwise, the maximal coordinate 
$[\mathbf{v}]_\ell$ of $\mathbf{v}$ must be strictly positive and unique since 
$\vv \notin \calK(\R^n)$. As a result, we can write 
$\zero_n = \vv - [\vv]_{\ell} \one_n 
+\sum\limits_{k \neq \ell}([\vv]_\ell-[\vv]_k)\ve_k$, and thus the zero vector 
can be expressed as a positive linear combination of the columns of $\mM_{\ell}$. 
Using Proposition~\ref{prop:characterizepss}, it follows that the matrix $\mM_{\ell}$ is a PSS.
\end{proof}

Our characterization of positive bases will consist in identifying critical 
matrices within a matrix decomposition, called the critical structure, that 
applies to any IN matrix.

\begin{definition}[Critical structure]~\label{def:critstruct}
Let $\mM \in \R^{n\times (n+s)}$ be an IN matrix as described in~(\ref{eq:pbIN}).  
There exist positive integers $n_1,\dots,n_s$ satisfying $n_1+\dots+n_s=n$ and 
matrices $\mX_1,\dots,\mX_{s-1}$ such that
\begin{equation}\label{eq:criticalstructure}
    \mM=\bbm \mM_1\\ \dots\\ \mM_s\ebm, 
    \quad \text{where}
    \left\{
    \begin{array}{ll}
        \mM_1=\bbm \mI_{n_1}&\mathbf{O}&-\one_{n_1}&\mX_1\ebm\\
        \mM_i=\bbm \mathbf{O}&\mI_{n_i}&\mathbf{O}&-\one_{n_i}&\mX_i\ebm
        &\text{if\ } i \in [\![2,s-1]\!] \\
        \mM_s=\bbm \mathbf{O}&\mI_{n_s}&-\one_{n_s}\ebm\\
    \end{array}
    \right.
\end{equation}
in which blocks of zeroes of arbitrary sizes are noted $\mathbf{O}$. The structure described by equation (\ref{eq:criticalstructure}) is called the \emph{critical structure} of $\mathbf{M}$.
\end{definition}

\rev{For instance, the matrix from Example~\ref{ex:PSSnetmatrix} has the critical 
structure 
\[
    \bbm \mM_{11} \\ \mM_{12} \ebm, 
    \quad \mbox{where} \quad 
    \mM_{11} = \bbm \mI_3 &\zero_{3,1} &-\one_3 &-\ve_3 \ebm, 
    \quad \mbox{and} \quad
    \mM_{12} = \bbm \zero_{1,3} &1 &0 &-1 \ebm.
\]}

When $s=1$, $\mM$ has the trivial structure $\mM=\mM_1=\bbm\mI_n&-\one_n\ebm$, which 
does not involve any arbitrary block $\mX_i$, and it is then immediate that $\mM$ 
is a positive basis. In the general case, however, assessing whether an IN matrix 
$\mM\in \R^{n\times(n+s)}$ with $\ell=n$ and $k=s\geq2$ is a positive basis amounts 
to checking the critical nature of the blocks $\mX_1,\dots,\mX_{s-1}$.

\begin{theorem}~\label{thm: M posbasis iff critical}
Let $\mM \in \R^{n\times(n+s)}$ be an IN matrix as described in (\ref{eq:pbIN}). 
Then, $\mM$ is a positive basis if and only if and each block $\mX_i$ in the 
critical structure~(\ref{eq:criticalstructure}) is a critical matrix in $\R^{n_i}$.
\end{theorem}
\begin{proof}
We first proceed by contrapositive and suppose that there exists 
$i \in [\![1,s-1]\!]$ such that $\mX_i$ is \emph{not} critical. Then, there exist 
$j_0 \in [\![1,n_i+1]\!]$ and $\mathbf{v} \in \pspan(\mX_i)$ such that the matrix 
obtained by replacing the $j_0^{th}$ column of 
$\bbm\mathbf{I_{n_i}}&-\one_{n_i}\ebm$ with $\mathbf{v}$ is a PSS. Let $j$ by the 
index of this column in $\mathbf{M_i}$, let $\mathbf{m_j}$ be the $j^{th}$ column of $\mathbf{M}$ and let $\mM'$ be the matrix obtained by removing $\mathbf{m_j}$ from $\mM$. Consider the matrix $\mM''=\bbm\mathbf{X_i}\\\mathbf{O}\ebm$ with 
$m'=n-m-n_i$ rows, where $m=n_1+\cdots+n_{i-1}$ so that $\mM''$ is a submatrix of $\mM'$. Using that $\vv \in \pspan(\mX_i)$, it follows that
\begin{equation}~\label{eq:Mu=xv0} 
    \mathbf{M'}\mathbf{u}=
    \bbm \mathbf{x}\\\mathbf{v}\\ \zero_{m'} \ebm 
    \quad \text{with} \quad 
    \mathbf{u}\geq \zero_{n+s-1}
    \quad \text{and} \quad \mathbf{x} \in \R^{m}.
\end{equation}
Moreover, there exists a PSS $\hat{\mM}$ of $\R^m$ and a zero matrix $\mathbf{O}$ such that $\bbm \hat{\mM} \\ \mathbf{O}\ebm$ is contained in $\mM'$. Therefore 
\rev{equation~\eqref{eq:Mu=xv0}} has a solution for any $\mathbf{x} \in \R^m$ and in particular, one can choose $\mathbf{u}$ such that $\mM'\mathbf{u}=\mathbf{m_j}$. In other words, $\mathbf{m_j}\in \pspan(\mathbf{M'})$, thus $\mathbf{M'}$ is a PSS and $\mathbf{M}$ is not a positive basis.

Conversely, suppose that all matrices $\{\mX_i\}_{i \in [\![1,s-1]\!]}$ are 
critical. Let $\vc$ be a column of $\mM$ and let 
$\mathbf{M'}=\bbm \mathbf{M'}_1 \\ \dots \\ \mathbf{M'}_s \ebm$ be the matrix 
obtained from $\mM$ by removing $\vc$. By construction of $\mM'$, there exists 
$i \in [\![1,s-1]\!]$ such that one of the columns of 
$\bbm\mI_{n_i}&-\one_{n_i}\ebm$ is not a column of $\mM'_i$. Let $\vv_i$ denote 
that column. By criticality of $\mX_i$, $\vv \notin \pspan(\mM'_i)$, and thus 
$\mM'_i$ is not a PSS. As a result, $\mM$ is no longer a PSS when one of its  
columns is removed, and thus it is a positive basis.
\end{proof}

It follows from Theorem~\ref{thm: M posbasis iff critical} that characterizing 
positive bases in $\R^n$ reduces to describing critical matrices in $\R^{m}$ for 
all $m\leq n$. The next proposition lists such matrices in dimension $\rev{n\leq 2}$, 
and will be used to characterize positive bases of near-maximal size.

\begin{proposition}~\label{prop: lowdim critical} 
\begin{enumerate}[(i)]
    \item $\mX \in \R^{1 \times m}$ is critical if and only if it is the zero matrix.
    \item $\mX \in \R^{2 \times m}$ is critical if and only if $\cols(\mX)$ is 
    contained in one of the cones defining $\calK(\R^2)$.
%    \item $\mX \in \R^{3 \times m}$ is critical if and only if $\cols(\mX)$ is 
%    contained in one of the cones defining $\calK(\R^3)$ or in 
%    $\spann(\ve_i+\ve_j)$ for some $(i,j) \in [\![1,3]\!]^2,\ i<j$.
\end{enumerate}
\end{proposition}
\begin{proof}
In all cases, the reverse implication is immediate, thus we focus on proving the 
direct implication. To this end, we recall that a critical matrix 
$\mX \in \R^{n \times m}$ satisfies 
$\cols(\mX) \subset \pspan(\mathbf{X}) \subset \mathcal{K}(\R^n)$ by 
definition.

\emph{(i)} Since $\calK(\R^1)=\{0\}$, the result is immediate.

\emph{(ii)} Consider any pair $\{\vx,\vy\} \subset \cols(\mX)$. If $\vx =\zero_2$, then $\vx$ and $\vy$ belong to one of the cones defining $\calK(\R^2)$. Now, suppose that $\vx \neq \zero_2$. If $\vx \in \calK_1(\R^2)$, then 
necessarily $[\vx]_2<0$. If $\vy \in \calK_2(\R^2)\setminus \calK_1(\R^2)$, then 
$\vx+\vy < \zero_2 \notin \calK(\R^2)$, which contradicts the criticality of 
$\mX$. Similarly, if $\vy \in \calK_{1,2}(\R^2)\setminus \calK_1(\R^2)$, then the 
vector $\tfrac{-[\vy]_1}{[\vx]_2}\vx+\vy$ is not critical since it is a positive 
multiple of $\ve_1$. Thus, $\vy \in \calK_1(\R^2)$. A similar reasoning for 
$\vx \in \calK_2(\R^2)$ and $\vx \in \calK_{1,2}(\R^2)$ shows that $\vx$ and $\vy$ 
must also lie in the same cone among those defining $\calK(\R^2)$. By the 
pidgeonhole principle, it follows that all vectors in $\cols(\mX)$ must lie in one 
cone among those defining $\calK(\R^2)$.
\end{proof}

\rev{Proposition~\ref{prop: lowdim critical} and Theorem~\ref{thm: M posbasis iff critical} can be 
used to characterize positive bases in $\R^n$ with critical 
structure~\eqref{eq:criticalstructure} consisting of blocks $\mM_i$ with at most 
two rows. When all blocks $\mM_i$ have one row, we recover the result of 
Theorem~\ref{th:structminmaxpb} since all $\mX_i$ blocks are necessarily zero 
matrices. Tackling the other two cases requires the following auxiliary result.}

\begin{lemma}~\label{lemma:blocks n_i=2 n_i=3 simplified}
Suppose $n \ge 2$. Any positive basis $\mathbf{D}_{n,s}$ is structurally equivalent 
to an IN matrix as described in~\eqref{eq:pbIN} whose critical 
structure~\eqref{eq:criticalstructure} satisfies 
$\cols(\mX_i)\subset \mathcal{K}_1(\R^2)$ whenever $n_i=2$.
%and $\cols(\mathbf{X}_i) \subset \mathcal{K}_{1}(\R^3)$ or 
%$\cols(\mathbf{X}_i) \subset \spann\left(\mathbf{e_2}+\mathbf{e_3}\right)$ 
%whenever $n_i=3$.
\end{lemma}
\begin{proof}
Let $\mM\equiv \mD_{n,s}$ be an IN matrix satisfying~(\ref{eq:pbIN}). 
Suppose that one of the blocks $\mM_i$ in its critical 
structure~\eqref{eq:criticalstructure} has two rows and let 
$m_i=n_1+\dots+n_{i-1}+1$ be the index of its first row in $\mM$. Since 
$\mD_{n,s}$ is a positive basis, the block $\mX_i$ is critical for $\R^2$ per 
Theorem~\ref{thm: M posbasis iff critical}. If 
$\cols(\mX_i) \subset \calK_1(\R^2)$ there is nothing to prove. Otherwise, we
must have $\cols(\mX_i) \subset \mathcal{K}_2(\R^2)$ or 
$\cols(\mX_i) \subset \mathcal{K}_{1,2}(\R^2)$ by 
Proposition~\ref{prop: lowdim critical}. If $\cols(\mX_i) \subset \calK_2(\R^2)$, 
permuting the rows and columns of indices $m_i$ and $m_i+1$ in $\mM$ (as well as 
other columns if needed) creates a new matrix $\mM'$ whose critical 
structure~\eqref{eq:criticalstructure} satisfies 
$\cols(\mX'_i) \subset \mathcal{K}_1(\R^2)$ and $\mX'_j = \mX_j$ whenever 
$j \neq i$. If $\cols(\mathbf{X}_i)\subset\mathcal{K}_{1,2}(\R^2)$, replacing the 
$(m_i+1)^{th}$ row of $\mM$ by its opposite and adding this new row to that of 
index $m_i$ creates again a matrix $\mM'$ with the desired critical structure up 
to further column permutation.
%
%Suppose now that the block $\mM_i$ has three rows. Since $\mX_i$ must be critical 
%for $\R^3$, its columns must lie in one of the cones stated in 
%Proposition~\ref{prop: lowdim critical}. If 
%$\cols(\mathbf{X}_i) \subset \spann(\mathbf{e_2}+\mathbf{e_3})$ or 
%$\cols(\mathbf{X}_i) \subset \mathcal{K}_1(\R^3)$, the desired conclusion follows. 
%More generally, if $\cols(\mX_i) \subset \spann(\mathbf{e_j}+\mathbf{e_k})$ or 
%$\cols(\mX_i) \subset \mathcal{K}_j(\R^3)$ for any $(j,k)$ in $[\![1,3]\!]$, 
%permutations of rows and columns can be applied to $\mM$ to obtain a matrix 
%$\mM'$ whose critical structure~\eqref{eq:criticalstructure} satisfies 
%$\cols(\mX_i')\subset \calK_1(\R^3)$ or 
%$\cols(\mX_i')\subset \spann(\ve_2+\ve_3)$ as well as $\mX'_j=\mX_j$ whenever 
%$j\neq i$. By Proposition~\ref{prop: lowdim critical} and without loss of 
%generality, the only remaining case to consider is that where 
%$\cols(\mX_i) \subset \mathcal{K}_{1,2}(\R^3)$. In that case, up to column 
%permutations, the matrix $\mM'$ described above can be obtained by replacing the 
%$m_i^{th}$ row of $\mM$ by its opposite and then adding it to the rows of index 
%$(m_i+1)$ and $(m_i+2)$.
\end{proof}

We are now equipped with the necessary tools to extend the result of 
Theorem~\ref{th:structminmaxpb} to other sizes of positive bases.

%%%%%%%%%%%%%%%%%%%%%%%%%%%%%%%%%%%%%%%%%%%%%%%%%%%%%%%%%%%%%%%%%%%%%%%%%%%%%%%
\subsection{Structure of nearly extreme-size positive bases}
\label{ssec:structnearextremepb}

Having shown in Section~\ref{ssec:crit} how the structure of any positive 
basis connects to the critical structure of its associated IN matrix, we now 
apply these results to fully characterize positive bases of size $2\ell-1$ and 
$\ell+2$ of an $\ell$-dimensional subspace of $\R^n$. To our 
knowledge, those descriptions are novel in both the positive spanning set and the 
minimal strongly edge-connected digraph literature. For this reason, we state every 
result using positive bases then provide an equivalent for strongly edge-connected 
digraphs, starting with the size $2\ell-1$.

%Our first result tackles positive bases of size $2n-1$.
\begin{theorem}~\label{th: struct bases 2n-1}
Let $n\geq \ell \geq 2$. A matrix 
$\mathbf{D}_{\LL,\ell-1} \in \R^{n \times (2\ell-1)}$ is a positive basis for 
some $\ell$-dimensional linear space $\LL \subset \R^n$ if and only if there exists 
a non-positive vector $\mathbf{x} \in \R^{\ell-2}$ such that
\begin{equation}
\label{eq:structpb2n1}
    \mathbf{D}_{\LL,\ell-1}
    \equiv
    \bbm \mathbf{I_{\ell}}&\mathbf{N}\\ 
    \zero_{n-\ell,\ell}&\zero_{n-\ell,\ell-1}\ebm, 
    \quad \text{with} \quad 
    \mathbf{N}=
    \bbm -1& \zero_{\ell-2}^\top \\ -1& \mathbf{x}^\top \\ 
    \zero_{\ell-2}&-\mathbf{I_{\ell-2}}\ebm.
\end{equation}
\end{theorem}
\begin{proof}
We only prove the result when $\ell=n$ as the reasoning easily adapts to all other 
cases. If $\ell=n$ and $\mD_{\LL,\ell-1}$ satisfies~\eqref{eq:structpb2n1}, the 
result holds. Conversely, let $\mD_{n,n-1}$ be a positive basis and 
$\mM=\bbm \mI_n&\mN\ebm$ be the associated IN matrix through~\eqref{eq:pbIN}. 
Since $\mN \in \R^{n\times (n-1)}$, the critical 
structure~(\ref{eq:criticalstructure}) of $\mM$ must consist of $n-1$ blocks 
$\mM_i \in \R^{n_i\times(2n-1)}$. Using $n_1+\dots+n_{n-1}=n$, it follows that 
exactly one of the blocks $\mM_i$ has two rows while the others have one. By 
Theorem~\ref{thm: M posbasis iff critical} and 
Proposition~\ref{prop: lowdim critical} each block $\mX_i$ must be critical and 
those with one row are zero matrices. Finally, 
Lemma~\ref{lemma:blocks n_i=2 n_i=3 simplified} implies that the unique block 
$\mX_i$ on two rows can be transformed into a matrix 
$\bbm 0_{\ell-2}^\top\\\vx^\top\ebm$ where $\vx \in \R^{\ell-2}$ is 
non-positive, hence the conclusion.
\end{proof}

An implication of Theorem~\ref{th: struct bases 2n-1} in terms of strongly 
edge-connected digraphs is given below, and illustrated through 
Example~\ref{ex:graph2n1}.

\begin{corollary}~\label{cor:mingraph2n-3}
A digraph $G=(V,A)$ on $n$ vertices and $2n-3$ arcs is minimally strongly 
edge-connected if and only if it is the union of a bi-directed forest $F$ with a 
graph $G\backslash F$ such that
\begin{enumerate}[(i)]
    \item $G\backslash F$ consists of circuits of size $3$, all sharing a common 
    arc $a \in A$.
    \item Each tree in $F$ contains exactly one vertex of $G\backslash F$.
\end{enumerate}
\end{corollary}
\begin{proof}
Suppose that $G$ is minimally strongly edge-connected and let $\mM=\mD_{n-1,n-2}$ in 
$\R^{(n-1)\times(2n-3)}$ be an associated network matrix. By 
Theorem~\ref{th: struct bases 2n-1}, 
$\mM \equiv \tilde{\mM}=\bbm \mI_{n-1}&\mN\ebm$, with 
$\mN=\bbm -1& \zero_{n-3}^\top \\ -1& \vx^\top \\ \zero_{n-3}&-\mI_{n-3}\ebm$ 
and $\vx \leq \zero_{n-3}$. By Lemma~\ref{lem:nicenetworkmatrix}, $\tilde{\mM}$ can 
be assumed to be a network matrix, implying $\vx \in \{-1,0\}^{n-3}$. 
Write $A=\{a_1,\dots,a_n\}$ so that the $i^{th}$ column of $\tilde{\mM}$ is 
associated to $a_i,$ for all $i$. 
Recalling that $\{a_{i_1},\dots,a_{i_k}\} \subset A$ is a circuit in $G$ if and only 
if it is associated to an inclusion-wise minimal set of columns in $\tilde{\mM}$ 
which add up to $\zero_{n-1}$. Based on this remark, we observe that 
$\{a_1,a_2,a_n\}$ is a circuit of size $3$ in $G$. Moreover, any other circuit of 
this size in $G$ is given by $\{a_2,a_{i+2},a_{n+i}\}$ where $i \in [\![1,n-3]\!]$ 
satisfies $[\vx]_i=-1$. We denote by $G'=(V',A')$ the union of these circuits and we 
let $|A'|=m$. Now, consider the spanning tree $T=(V,\hat{A})$ associated to 
$\tilde{\mM}$. $T$ can be decomposed as the union of $T\cap G'$ with a sequence 
$(T_j)_{j\leq m}$ of pairwise disjoint trees, each touching a unique vertex in $G'$. 
It is easy to see that each arc in any tree $T_j$ is part of a circuit of size $2$. 
Indeed, any such arc $a_i$ must satisfy $i \in [\![3,n-1]\!]$ and 
$[\vx]_{i-2}=0$, and as such is contained in a circuit $\{a_i,a_{n+i-2}\}$. We have 
thus established (i) and (ii).

Conversely, if a graph $G$ admits such a decomposition, it must be edge-connected 
and $\tilde{\mM}$ is a network matrix for $G$. Since $\tilde{\mM}$ is a positive 
basis, it follows that $G$ is minimally strongly edge-connected. 
\end{proof}
\begin{example} 
\label{ex:graph2n1}
A minimally strongly edge-connected digraph on $9$ vertices and $15$ 
arcs, where $F$ is in black and $G\backslash F$ is in blue.
\begin{figure}[H]
\centering
\begin{tikzpicture}
\tikzstyle{vertex}=[circle, fill=white, minimum size=5pt, inner sep=1pt]
\tikzset{->-/.style={decoration={
  markings,
  mark=at position .5 with {\arrow{>}}},postaction={decorate}}}
\tikzstyle{edge}=[->-,thick]
\node[vertex,blue] (u) at (0,0) {};
\node[vertex,blue] (v) at (4,0) {};
\node[vertex,blue] (w_1) at (2,-1) {};
\node[vertex,blue] (w_2) at (2,-1.5) {};
\node[vertex,blue] (w_3) at (2,-2) {};
\node[vertex,fill=none] at (1.95,0.3) {$a$};
%\node[vertex,fill=none] at (0,0.3) {$u$};
%\node[vertex,fill=none] at (4,0.3) {$v$};

\draw[edge,blue] (u) to (v);
\draw[edge,blue] (v) to (w_1);
\draw[edge,blue] (w_1) to (u);
\draw[edge,blue] (v) to (w_2);
\draw[edge,blue] (w_2) to (u);
\draw[edge,blue] (v) to (w_3);
\draw[edge,blue] (w_3) to (u);
\node[vertex,black] (t_1) at (-1,0) {};
\node[vertex,black] (t_2) at (-2,1) {};
\node[vertex,black] (t_3) at (-2,-1) {};
\draw[edge,bend right=30] (u) to (t_1);
\draw[edge,bend right=30] (t_1) to (u);
\draw[edge,bend right=30] (t_1) to (t_2);
\draw[edge,bend right=30] (t_2) to (t_1);
\draw[edge,bend right=30] (t_1) to (t_3);
\draw[edge,bend right=30] (t_3) to (t_1);
\node[vertex,black] (t_4) at (5,0) {};
\draw[edge,bend right=30] (v) to (t_4);
\draw[edge,bend right=30] (t_4) to (v);

\end{tikzpicture}
\end{figure}

\end{example}

We now turn to positive bases of size $\ell+2$ and the associated strongly 
edge-connected digraphs.

\begin{theorem}~\label{th: struct pb n+2}
Let $n\geq \ell \geq 2$. A matrix $\mD_{\LL,2} \in \R^{n\times(\ell+2)}$ is a positive basis for some $\ell$-dimensional linear space $\LL \subset \R^n$ if and only if there exists a non-positive vector $\mathbf{x}\in \R^k$, $k \in [\![1,\ell-1]\!]$ satisfying $[\mathbf{x}]_1=0$ such that
$$\mathbf{D}_{\LL,2}\equiv\bbm \mathbf{I_\ell}&\mathbf{N}\\\zero_{n-\ell,\ell}&\zero_{n-\ell,2}\ebm \quad \text{with} \quad \mathbf{N}=\bbm -\one_k&\mathbf{x} \\ \zero_{\ell-k}&-\one_{\ell-k}\ebm.$$
\end{theorem}
\begin{proof}
As for Theorem~\ref{th: struct bases 2n-1} , the converse implication is trivial and will not be proved. Moreover, we assume that $\ell=n$ as the general result is easily deduced from this specific case. Consider a positive basis $\mathbf{D}_{n,2}$ of size $n+2$ in $\R^n$ and any IN matrix $\mM=\bbm\mI_n&\mN\ebm$ associated to it through~(\ref{eq:pbIN}) and note $\mathbf{N}=\bbm -\one_k&\mathbf{w}\\\zero_{n-k}&-\one_{n-k}\ebm$ where $k\in[\![1,n-1]\!]$ and $\mathbf{w} \in \R^k$ is arbitrary. By Theorem~\ref{thm: M posbasis iff critical}, $\mathbf{w}$ is critical: if this vector is non-positive, one of its coordinates must be zero. In this case, the result is proved after permuting rows and columns in $\mM$.\\
Otherwise and by definition of a critical vector, the maximal entry of $\mathbf{w}$ must be non-unique and positive. Let $[\mathbf{w}]_i=[\mathbf{w}]_j>0$ be two maximal entries and denote by $\mathbf{u}$ and $\mathbf{v}$ the two columns of $\mN$. One easily checks that $\mathcal{B}=\{\mathbf{u},\mathbf{e_{1}}\dots,\mathbf{e_{n}}\}\backslash\{\mathbf{e_{i}}\}$ is a linear basis for $\R^n$, moreover
$$\mathbf{e_{i}}=-\mathbf{u}-\sum\limits_{\ell=1, \ell \neq i}^{k}\mathbf{e_\ell} \quad \text{and} \quad \mathbf{v}=-[\mathbf{v}]_i\mathbf{u}+\sum\limits_{\ell=1, \ell \neq i}^{k}([\mathbf{v}]_\ell-[\mathbf{v}]_i)\mathbf{e_\ell}-\sum\limits_{\ell=k+1}^{n} \mathbf{e_\ell}.$$
Note that each coefficient in the linear combination associated to $\mathbf{e_{i}}$ is non-positive. Similarly, the linear combination associated to $\mathbf{v}$ is non-positive as $[\mathbf{v}]_i=[\mathbf{w}]_i$ is a maximal entry of $\mathbf{v}$, and this combination is not strictly negative as $[\mathbf{v}]_j-[\mathbf{v}]_i=0$. 
In consequence, letting $\mathbf{B} \in \R^{n\times n}$ satisfy $\cols(\mathbf{B})=\mathcal{B}$, we see, up to rows and columns permutation, \rev{that} the matrix  $\mathbf{B}^{-1}\mM$ has the announced structure. Since $\mD_{n,2}\equiv\mathbf{B}^{-1}\mM,$ the result is proved.
\end{proof}

%As was the case earlier, the characterization given in Theorem~\ref{th: struct pb n+2} can be applied to network matrices to derive a characterization on strongly connected digraphs.
\begin{corollary}~\label{cor: mingraph n+1}\
A digraph $G=(V,A)$ on $n$ vertices and $n+1$ arcs is minimally strongly edge-connected if and only if it is the union of two circuits whose intersection defines an elementary path - potentially reduced to a single vertex - in $G$.
\end{corollary}%
\begin{proof}
We only prove the direct implication as the converse is trivial. Let $\mM=\mD_{n-1,2}$ in $\R^{(n-1)\times(n+1)}$ be a network matrix associated to the minimally strongly connected graph $G$. By Theorem~\ref{th: struct pb n+2}, $\mM \equiv \tilde{\mM}$ where 
$$\tilde{\mM}=\bbm \mathbf{I}_{n-1}&\mathbf{N}\ebm, 
\quad \text{with} \quad 
\mathbf{N}=\bbm -\one_k& \mathbf{x} \\ \zero_{n-1-k}&-\one_{n-1-k}\ebm 
\quad \text{and} \quad \mathbf{x}\leq \zero_{k},\quad k\geq 1, \quad [\mathbf{x}]_1\neq 0.$$ 
Using Lemma~\ref{lem:nicenetworkmatrix} we assume $\tilde{\mM}$ to be a network matrix, therefore  $\mathbf{x} \in \{-1,0\}^{k}$. Write $A=\{a_1,\dots,a_n\}$ and suppose that the $i^{th}$ column of $\tilde{\mM}$ is associated to $a_i,$ for all $i$.\\
    Based on the structure of $\tilde{\mM}$ the graph contains a circuit of size $k+1$, namely $\mathcal{C}_1=\{a_1,\dots,a_k,a_n\}$. Moreover $\mathcal{C}_2=\{a_{i_1},\dots,a_{i_m},a_{k+1},\dots,a_{n-1},a_{n+1}\}$ is a circuit in $G$, where $\{i_1,\dots,i_m\}\subset [\![2,k]\!]$ is the - potentially empty - set of indices $i$ such that $[\mathbf{x}]_i=-1$. We show that $\mathcal{C}_1\cap\mathcal{C}_2$ is an elementary path. For $i \leq 2$, let $V_i\subset V$ be the set of extremities of the arcs in $\mathcal{C}_i$ and let $v \in V_1$. Consider the sequence given by $G_0=(\{v\},\emptyset)$, $G_1=(V_1,\mathcal{C}_1)$, $G_2=(V_1\cup V_2,\mathcal{C}_1\cup\mathcal{C}_2)$. As $\mathcal{C}_1\cup\mathcal{C}_2=A$, the sequence defines an ear-decomposition for $G$. In particular $(\mathcal{C}_1\cup\mathcal{C}_2)\backslash\mathcal{C}_1=\mathcal{C}_2\backslash\mathcal{C}_1$ is an ear of $G$ that defines a $u-v$ path between two vertices in $V$. Then, since $\mathcal{C}_2$ is a circuit, we find that $(\mathcal{C}_1\cup\mathcal{C}_2)\cap\mathcal{C}_1=\mathcal{C}_1\cap\mathcal{C}_2$ must be an elementary $v-u$ path.
\end{proof}

Theorems~\ref{th:structminmaxpb},~\ref{th: struct bases 2n-1} 
and~\ref{th: struct pb n+2} together imply that any positive basis 
$\mD_{n,s}, n\leq 4$ is associated through~\eqref{eq:pbIN} to an IN matrix 
$\bbm \mI_n&\mN\ebm$ where $\mN$ has non-positive entries. 
%Accordingly, any minimally strongly edge-connected digraph $G=(V,A)$ on $5$ 
%vertices or less contains a spanning tree $T=(V,\hat{A})$ in which any $u-v$ path 
%associated to an arc of $A\backslash \hat{A}$ exclusively uses arcs in the 
%backward direction. 
%
Although positive bases can always be generated from such IN matrix 
structures~\cite[Theorem 5.4]{RGRegis_2016}, we emphasize that the equivalence 
no longer holds in dimension $5$ or higher. Indeed, the matrix 
$
\mathbf{D}_{5,8}
=
\bbm \mathbf{I_5}
&-\mathbf{e_1}-\mathbf{e_2}-\mathbf{e_3}
&-\mathbf{e_2}-\mathbf{e_3}-\mathbf{e_4}
&\mathbf{e_2}+\mathbf{e_3}-\mathbf{e_5}
\ebm
$ 
is both a positive basis and a network matrix, but one it cannot be associated 
through~\eqref{eq:pbIN} to an IN matrix with a non-positive $\mathbf{N}$ block.

%%%%%%%%%%%%%%%%%%%%%%%%%%%%%%%%%%%%%%%%%%%%%%%%%%%%%%%%%%%%%%%%%%%%%%%%%%%%%%%
\section{Conclusion}
\label{sec:conc}
%%%%%%%%%%%%%%%%%%%%%%%%%%%%%%%%%%%%%%%%%%%%%%%%%%%%%%%%%%%%%%%%%%%%%%%%%%%%%%%

We have introduced a matrix decomposition technique inspired by the ear 
decomposition for strongly connected digraphs, that can be used as a 
certificate for assessing the positive spanning nature of a matrix. Our study 
also sheds a new light on the relationship between PSSs and strongly connected 
digraphs, the latter giving rise to network matrices of the former nature.

Our study can be extended in a number of research directions. A natural 
continuation of the present work consists in adapting our results to 
orthogonally structured positive bases or positive $k$-spanning 
sets~\cite{WHare_GJarryBolduc_SKerleau_CWRoyer_2024}. Exploiting our 
decomposition in the context of optimization algorithms is also a future area 
of investigation.

\paragraph{Acknowledgments} We thank the guest editors as well as two 
anonymous referees for their comments.

%%%%%%%%%%%%%%%%%%%%%%%%%%%%%%%%%%%%%%%%%%%%%%%%%%%%%%%%%%%%%%%%%%%%%%%%%%%%%%%
\bibliographystyle{siam}
\bibliography{refsPB}
%%%%%%%%%%%%%%%%%%%%%%%%%%%%%%%%%%%%%%%%%%%%%%%%%%%%%%%%%%%%%%%%%%%%%%%%%%%%%%%

%%%%%%%%%%%%%%%%%%%%%%%%%%%%%%%%%%%%%%%%%%%%%%%%%%%%%%%%%%%%%%%%%%%%%%%%%%%%%%%
%%%%%%%%%%%%%%%%%%%%%%%%%%%%%%%%%%%%%%%%%%%%%%%%%%%%%%%%%%%%%%%%%%%%%%%%%%%%%%%
%%%%%%%%%%%%%%%%%%%%%%%%%%%%%%%%%%%%%%%%%%%%%%%%%%%%%%%%%%%%%%%%%%%%%%%%%%%%%%%
\end{document}